\documentclass{article} 
\usepackage{iclr2026_conference,times}

\definecolor{highlightyellow}{RGB}{255,255,204}
\definecolor{leftbarcolor}{RGB}{255,204,0}

\newcommand{\eg}{\hbox{\emph{e.g.,}}\xspace}
\newcommand{\ie}{\hbox{\emph{i.e.,}}\xspace}

\usepackage{xcolor}

\definecolor{darkblue}{RGB}{0,0,139} 

\usepackage[colorlinks=true,
            citecolor=darkblue,
            linkcolor=black,
            urlcolor=darkblue]{hyperref}

\usepackage{url}
\usepackage{wrapfig}
\usepackage[dvipsnames]{xcolor} 
\usepackage{tabularx}
\usepackage[most]{tcolorbox} 
\usepackage{colortbl}
\usepackage{tabu}
\usepackage{booktabs}
\usepackage{multirow}

\usepackage[T1]{fontenc}
\usepackage[utf8]{inputenc}

\usepackage{graphicx} 
\usepackage{float}
\usepackage{subfigure} 
\usepackage{amssymb}

\usepackage{tablefootnote}
\usepackage{xspace}

\usepackage{float}
\usepackage{amsmath}
\usepackage{cleveref}
\usepackage{bm}
\usepackage{algorithmicx}
\usepackage{algorithm}
\usepackage{algpseudocode}

\usepackage{arydshln}

\usepackage{amssymb}
\usepackage{pifont}
\newcommand{\cmark}{\textcolor{green!35!black}{\ding{51}}}
\newcommand{\xmark}{\ding{55}}%
\usepackage{enumitem}
\usepackage{bm}

\usepackage{cuted}

\usepackage{gradient-text}
\usepackage{fontawesome5}

\usepackage{caption}
\usepackage{array}
\usepackage[subfigure]{tocloft}
\usepackage{makecell}

\usepackage{fancyhdr}

\usepackage{listings}
\addtocontents{toc}{\protect\setcounter{tocdepth}{-1}}

\renewcommand{\cite}{\citep}
\usepackage{multicol}

\usepackage{adjustbox}

\newcommand{\mrmr}{MRMR}
\newcommand{\ours}{\gradientRGB{\mrmr}{132, 60, 180}{0, 0, 0}\xspace}

\newcommand{\pin}{PIN-14M\xspace}
\newcommand{\mmmu}{MMMU\xspace}
\newcommand{\mteb}{MTEB\xspace}
\newcommand{\designqa}{DesignQA\xspace}
\newcommand{\mmmupro}{MMMU-Pro\xspace}
\newcommand{\bright}{\textsc{Bright}\xspace}

\newcommand{\tiir}{TIIR\xspace}
\newcommand{\mmrag}{MM-RAG\xspace}
\newcommand{\knowledge}{\emph{Knowledge}\xspace}
\newcommand{\theorem}{\emph{Theorem}\xspace}
\newcommand{\contradiction}{\emph{Contradiction}\xspace}
\newcommand{\gptsearch}{GPT-Search\xspace}

\newcommand{\totalnum}{1,435\xspace}

\title{\textbf{\ours}: A Realistic and Expert-Level Multidisciplinary Benchmark for Reasoning-Intensive Multimodal Retrieval}

\definecolor{YaleYellow}{RGB}{179, 176, 4} 
\definecolor{NYUPurple}{RGB}{134, 1, 175}  
\definecolor{NTUBlue}{RGB}{2,2,200} 
\definecolor{J}{RGB}{255, 106, 0}
\definecolor{Center}{RGB}{0, 128, 0}
\definecolor{UCASRed}{RGB}{150, 0, 0}      

\author{
\bf{Siyue Zhang$^{*\hspace{0.02em}{\textcolor{UCASRed}{\boldsymbol{N}}}}$$^{\hspace{.02em}\textcolor{UCASRed}{\boldsymbol{S}}}$ 
~~Yuan Gao$^{*\hspace{0.02em}{\textcolor{UCASRed}{\boldsymbol{J}}}}$
~~Xiao Zhou$^{*\hspace{0.02em}{\textcolor{UCASRed}{\boldsymbol{J}}}}$
~~Yilun Zhao$^{\hspace{.1em}\textcolor{UCASRed}{\boldsymbol{Y}}}$ 
~~ Tingyu Song$^{\hspace{.1em}\textcolor{UCASRed}{\boldsymbol{A}}}$
}
\\
\bf{
Arman Cohan$^{\hspace{.1em}\textcolor{UCASRed} {\boldsymbol{Y}}}$
~ Anh Tuan Luu$^{\hspace{0.02em}{\textcolor{UCASRed}{\boldsymbol{N}}}}$$^{\hspace{.02em}\textcolor{UCASRed}{\boldsymbol{V}}}$ \quad
~ Chen Zhao$^{\hspace{0.02em}{\textcolor{UCASRed}{\boldsymbol{S}}}}$$^{\hspace{.02em}\textcolor{UCASRed}{\boldsymbol{C}}}$ 
}
\vspace{9pt}\\
$^{\textcolor{UCASRed}{\boldsymbol{N}}}$Nanyang Technological University \quad
$^{\textcolor{UCASRed}{\boldsymbol{Y}}}$Yale University \quad
$^{\textcolor{UCASRed}{\boldsymbol{S}}}$NYU Shanghai \quad
\\
$^{\textcolor{UCASRed}{\boldsymbol{J}}}$Shanghai Jiao Tong University  \quad
$^{\textcolor{UCASRed}{\boldsymbol{A}}}$University of the Chinese Academy of Sciences
\\
$^{\textcolor{UCASRed}{\boldsymbol{C}}}$Center for Data Science, New York University
\quad
$^{\textcolor{UCASRed}{\boldsymbol{V}}}$VinUniversity
}

\iclrfinalcopy 

\begin{document}

\maketitle

\begin{abstract}

We introduce \ours, the first expert-level multidisciplinary multimodal retrieval benchmark requiring intensive reasoning. \ours contains \totalnum queries spanning 23 domains, with positive documents carefully verified by human experts. Compared to prior benchmarks, \ours introduces three key advancements. First, it challenges retrieval systems across diverse areas of expertise, enabling fine-grained model comparison across domains. Second, queries are reasoning-intensive, with images requiring deeper interpretation such as diagnosing microscopic slides. We further introduce Contradiction Retrieval, a novel task requiring models to identify conflicting concepts. Finally, queries and documents are constructed as image–text interleaved sequences. Unlike earlier benchmarks restricted to single images or unimodal documents, \ours offers a realistic setting with multi-image queries and mixed-modality corpus documents. We conduct an extensive evaluation of 4 categories of multimodal retrieval systems and 14 frontier models on \ours. The text embedding model Qwen3-Embedding with LLM-generated image captions achieves the highest performance, highlighting substantial room for improving multimodal retrieval models. Although latest multimodal models such as Ops-MM-Embedding perform competitively on expert-domain queries, they fall short on reasoning-intensive tasks. We believe that \ours paves the way for advancing multimodal retrieval in more realistic and challenging scenarios. 
\footnote{Our data is available at \url{https://huggingface.co/datasets/MRMRbenchmark}.}

\begin{figure}[t!]
    \centering
    \includegraphics[width=1\textwidth]{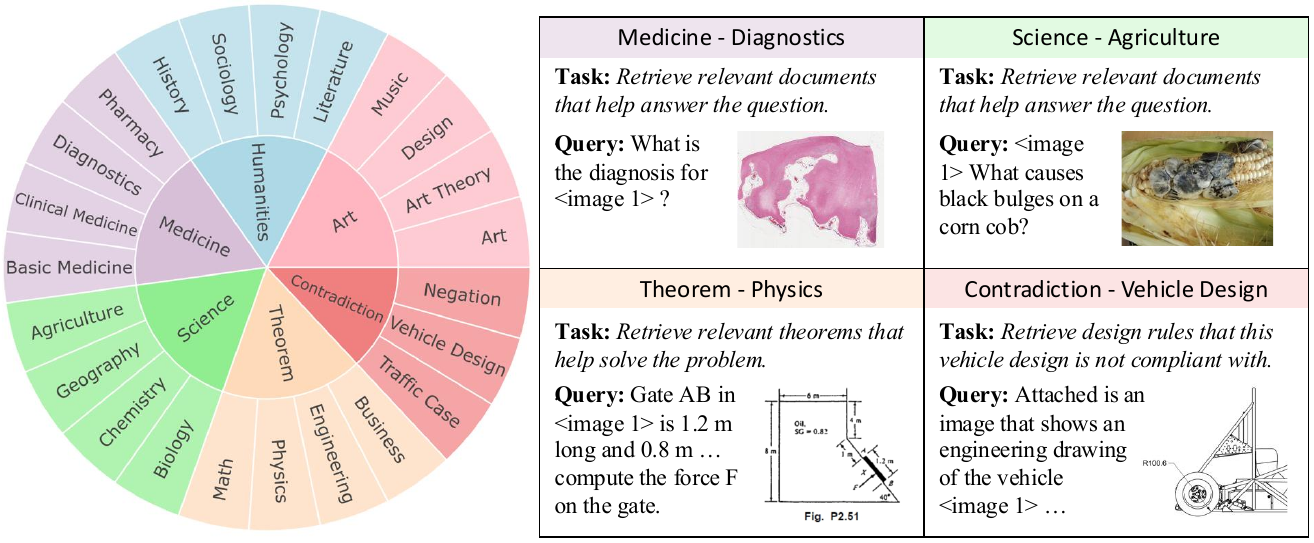}
    \captionsetup{justification=justified, singlelinecheck=false}  
    \caption{Overview of the \ours benchmark. \ours includes \totalnum expert-annotated examples, covering 23 domains across 6 disciplines. It is specifically designed to assess multimodal retrieval models in expert-level, reasoning-intensive tasks. Notably, we originally introduce the \emph{Contradiction Retrieval} task in the multimodal setting, which requires retrieving documents that conflict with the user query and features deeper logical reasoning.}
    \label{fig1}
\end{figure}
\end{abstract}

\section{Introduction}




LLM-based agents, such as DeepResearch~\cite{gptsearch, websearcher}, have been widely applied in domains including science, engineering, medicine, and finance \cite{medrag,medagents,finrag,humanitysexam}. These systems move beyond the intrinsic knowledge of LLMs by actively retrieving and integrating external information, making a strong and robust retrieval component essential~\cite{browsecomp-plus}. 
In practice, many expert-domain applications rely on multimodal information, underscoring the need for retrieval methods that can handle queries and documents spanning both visual and textual modalities, or even interleaved image–text content \cite{fashioniq, cirr, edis}.
For instance, given a medical image, the agent system should retrieve similar cases or guidelines to support clinical decisions.

While existing multimodal retrieval benchmarks have made progress, they are insufficient to capture the complexity of agentic scenarios. We identify three key limitations: 
(1) \textbf{Multidisciplinary expert domains}: most multimodal benchmarks are built on Wikipedia text and images, focusing on general-domain knowledge \cite{oven,infoseek,tiir}. However, state-of-the-art LLMs already demonstrate strong capabilities in handling such knowledge \cite{team2025kimi}, making it essential to develop benchmarks for high-stakes expert domains such as medicine, science, and engineering.
(2) \textbf{Reasoning intensity}: existing benchmarks primarily target semantic matching and information-seeking tasks, whereas real-world queries often involve expert-domain images and require deeper understanding and logical reasoning over them.
(3) \textbf{Image-text interleaving}: prior benchmarks mostly support single-image queries with supplementary text, yet real-world queries and documents typically consist of interleaved text and multiple images \cite{tiir}.

To address these gaps, we introduce \textbf{\ours}, a comprehensive benchmark measuring retrieval models in expert-level \textbf{\underline{M}}ultidisciplinary and \textbf{\underline{R}}easoning-intensive \textbf{\underline{M}}ultimodal \textbf{\underline{R}}etrieval. \autoref{fig1} presents an overview of our benchmark. \ours consists of \totalnum expert-annotated examples, categorized into three types of retrieval tasks: (1) \knowledge for retrieving web pages related to queries involving multiple expert-domain images; (2) \theorem for retrieving theorems involved in solving multimodal math problems; and (3) \contradiction for retrieving contradictory statements or rules given a case description. 
Specifically, we derive complex multidisciplinary queries from established Visual Question Answering (VQA) benchmarks \cite{mmmu,mmmupro} and assign expert annotators to collect positive documents from Internet.
To build a sizable corpus, we additionally include negative documents from knowledge-intensive collections \cite{pin,bright}. 
To further elevate the reasoning challenge, we originally introduce \emph{Contradiction Retrieval}, which requires models not only to detect semantic relevance but also to perform logical reasoning to identify conflicting concepts. 
To foster a deeper integration of visual and textual content, we represent both queries and documents in an interleaved multimodal format.

We conduct an extensive evaluation on \ours across four main categories of multimodal retrieval approaches and 14 representative models. The results reveal that current  multimodal retrieval systems consistently underperform text-only retrievers with image captioning on knowledge- and reasoning-intensive multimodal queries. The highest score of 54.1 is achieved by the text embedding model Qwen3-Embedding \cite{qwen3embedding} combined with LLM-based image captioning. The best-performing multimodal model, Ops-MM-Embedding \cite{ops}, trails by 6.0 points, mainly due to its limited reasoning capabilities rather than domain expertise. Its performance drops from 67.4 on \knowledge tasks to 37.4 and 36.6 on \theorem and \contradiction tasks, even though the corpora for these two tasks are much smaller than that of \knowledge. More importantly, the multidisciplinary setup in \ours reveals substantial performance differences across models and domains. For instance, Ops-MM-Embedding surpasses the second-best model, MM-Embed \cite{mmembed}, in the Art discipline, whereas their performances are comparable in the Medicine discipline. We hope our benchmark and findings will help progress in multimodal retrieval.

\section{Related Work}

\paragraph{Benchmarking multimodal retrieval.} As illustrated in \autoref{tab:data-comparison}, existing multimodal retrieval datasets mainly focus on semantic matching or information-seeking tasks. Early semantic matching benchmarks are built from paired image–text data, where the text is semantically aligned with the image \cite{edis,scimmir, mieb, vlm2vec}, and the task is to retrieve the corresponding modality. Composed Image Retrieval (CIR) emerges as a challenging task that allows users to search for target images using a multimodal query, comprising a reference image and a modification text specifying the user's desired changes to the reference image \cite{fashioniq, cirr, circo, magiclens}. Information-seeking benchmarks either retrieve supporting evidence for visual questions \cite{oven, infoseek} or retrieve multimodal documents for textual queries \cite{unifying, vidore, mmdocir}. As all prior studies focus on single-image inputs, \tiir \cite{tiir} proposes a more realistic setup in which the query and document consist of interleaved text–image sequences supporting multiple images. However, it is limited to searching general-domain wikiHow tutorials. To further advance multimodal retrieval, we construct \ours, the first benchmark comprising complex multidisciplinary queries that require in-depth reasoning in the interleaved text–image format.

\paragraph{Multimodal retrieval models and multimodal retrieval augmented generation.} State-of-the-art multimodal retrieval models commonly rely on large pre-trained encoders such as CLIP \cite{clip} and BLIP \cite{blip2}, which map images and texts into a shared embedding space. Their outputs are often combined using fusion strategies (\eg score fusion) to integrate information across modalities \cite{uniir}. More recent works adapt multimodal large language models (MLLMs) for universal multimodal embeddings by fine-tuning them on diverse retrieval tasks \cite{e5v, gme, vlm2vec, mmembed}. In these approaches, multimodal queries are processed through the MLLM, and the hidden states from the final transformer layer, typically the last token representation, are used as the dense embedding for retrieval. In this work, we benchmark a diverse set of multimodal retrieval approaches, including text retrievers with image captioning, text and image two-stream models with vector fusion, and multimodal retrievers. Additionally, thanks to advances in both retriever and generative models, multimodal retrieval-augmented generation (\mmrag) has emerged as a key application \cite{mragbench, mmsearch, mmsearch_r1, mmrag, realmmragrealworldmultimodalretrieval}. While various \mmrag benchmarks have been introduced, most focus on evaluating response generation and lack evidence-level relevance annotations, making it impractical to assess retrieval performance and its contribution within \mmrag \cite{browsecomp-plus}.

\begin{table}[t!]
\centering
\caption{Comparison of multimodal retrieval benchmarks and \ours. In the ``Modality'' column, ``T $\rightarrow$ I'' indicates retrieving image documents using a text query. The ``\#Domain'' column reports the number of domains; ``Open'' denotes datasets built from Wikidata with general domains. The ``Expert'', ``Reason'', and ``Interleaved'' columns indicate whether expert knowledge is required, whether intensive reasoning is involved, and whether data are in the interleaved image-text format. } 
\label{tab:data-comparison}
\begin{adjustbox}{width=\textwidth,center}
\begin{tabular}{l l l c c c c c}
\toprule
\textbf{Benchmarks} & \textbf{Modality} & \textbf{Retrieval Type} & \textbf{\#Domain} & \textbf{\#Query} & \textbf{Expert?} & \textbf{Reason?} & \textbf{Interleaved?} \\
\midrule

NIGHTS       & I $\rightarrow$ I &    Visual Similarity    & Open     &  20K   & \xmark              &         \xmark      &             \xmark     \\

SciMMIR         & T $\leftrightarrow$ I & Image Caption      & 11     & 530K      & \cmark              &         \xmark      &             \xmark     \\

EDIS & T $\rightarrow$ IT & Image Caption      & Open     & 3,241     & \xmark              &         \xmark      &             \xmark   \\
Wiki-SS         & T $\rightarrow$ I            & Document QA    & Open   & 3,610     &       \xmark   &            \xmark   &        \xmark     \\
WebQA        & T $\rightarrow$ IT            & Document QA  & Open   & 2,511     &       \xmark   &            \xmark   &        \xmark     \\
ViDoRe          & T $\rightarrow$ IT                               & Document QA     & 10     & 3,810     & \cmark              &        \xmark    &      \xmark     \\
MMDocIR         & T $\rightarrow$ IT                               & Document QA  & 10     & 1,658     & \cmark              &      \xmark        &       \xmark    \\
FashionIQ       & IT $\rightarrow$ I                               & Composed Image & 1    & 12,238    &    \xmark       &          \xmark     &       \xmark     \\
CIRR            & IT $\rightarrow$ I                               & Composed Image & Open & 4,148     &     \xmark       &         \xmark   &       \xmark      \\
CIRCO            & IT $\rightarrow$ I                               & Composed Image & Open & 1,020     &     \xmark       &         \xmark   &       \xmark      \\
InfoSeek        & IT $\rightarrow$ IT                              & VQA  & Open   & 1.35M     &        \xmark       &        \xmark       &        \xmark \\
OVEN            & IT $\rightarrow$ IT                              & VQA   & Open   & 18,341    &       \xmark     &  \xmark      &        \xmark     \\
wikiHow-TIIR    & IT $\rightarrow$ IT                              & VQA  & Open   & 7,654     &           \xmark    &       \xmark        & \cmark              \\
\midrule
\ours           & IT $\rightarrow$ IT                              & VQA & 23     &     \totalnum      & \cmark              & \cmark              & \cmark              \\
\bottomrule
\end{tabular}
\end{adjustbox}
\end{table}

\paragraph{Reasoning-intensive retrieval.} Beyond keyword- and semantic-based information retrieval, \bright \cite{bright} has introduced the first benchmark in the text domain that requires intensive reasoning to identify relevant documents. For example, given a new math or physics problem, the retrieval system is expected to provide previously solved problems using the same theorems or relevant theorem statements. To tackle this challenge, recent methods train the text retrievers using synthetic datasets containing complex queries and hard negatives \cite{rank1, rader, diffembed, diver, reasonir, bge_reasoner}. Our work extends reasoning-intensive retrieval into the multimodal domain. \ours\ is constructed by sourcing expert-level queries from the multimodal understanding and reasoning benchmark \mmmupro \cite{mmmupro}, collecting image-text interleaved documents from web pages, and obtaining relevance annotations from human experts.

\section{\ours Benchmark}
\label{sec: constructing ours}

\begin{table}[t]
\centering
\caption{Data statistics of \ours. For each dataset, we show the number of queries ($Q$) and documents ($D$), the average number of positive documents ($D_+$) per example, the average number of text tokens of queries and documents (measured by the GPT-2 tokenizer \cite{gpt2}, not including task instruction text), the average number of images in queries and documents, and sources of queries and documents. \knowledge datasets share a common retrieval corpus, while \theorem datasets share another. Examples for each dataset can be found in \autoref{examples}.}
\label{tab:data_stats}
\begin{adjustbox}{width=\textwidth,center}
\begin{tabular}{lcccccccccc}
\toprule
 & \multicolumn{3}{c}{Total Number} & \multicolumn{2}{c}{Avg. \#Text} & \multicolumn{2}{c}{Avg. \#Images} & \multicolumn{2}{c}{Source} & Ex. \\
\cmidrule(lr){2-4} \cmidrule(lr){5-6} \cmidrule(lr){7-8} \cmidrule(lr){9-10}
\textbf{Dataset} & $Q$ & $D$ & $D_+$ & $ Q $ & $ D $ & $Q$ & $D$ & $Q$ & $D$ & \\
\midrule
\multicolumn{11}{c}{\textit{Knowledge}} \\
\midrule
Art & 157 & 26,223 & 1.8 & 15.4 & 421.6 & 1.1 & 0.72 & \multirow{4}{*}{\makecell{\mmmupro \\ knowledge \\ question}} & \multirow{4}{*}{\makecell{\pin,\\Web pages}} & Fig.~\ref{case: music} \\
Medicine & 167 & 26,223 & 2.2 & 32.0 & 421.6 & 1.1 & 0.72 & & & Fig.~\ref{case: clinic} \\
Science & 137 & 26,223 & 1.8 & 32.1 & 421.6 & 1.2 & 0.72 & & & Fig.~\ref{case: biology} \\
Humanities & 94 & 26,223 & 1.9 & 54.5 & 421.6 & 1.2 & 0.72 & & & Fig.~\ref{case: psychology} \\
\midrule
\multicolumn{11}{c}{\textit{Theorem}} \\
\midrule
Math & 60 &  14,257 & 2.1 & 64.6 & 364.3  & 1.1 & 0.001 & \multirow{4}{*}{\makecell{\mmmupro \\ calculation \\ question}} & \multirow{4}{*}{\makecell{\bright, \\ Web pages}} & Fig.\ref{case: math} \\
Physics &  104 & 14,257  & 2.1 & 56.2 & 364.3 & 1.0 &  0.001& & & Fig.\ref{case: physics} \\
Engineering & 190  & 14,257  & 2.0 & 53.5 & 364.3 & 1.0 &  0.001 &  & & Fig.\ref{case: engineer} \\
Business & 158  & 14,257  & 3.2 & 64.2 & 364.3 & 1.0 &  0.001&   & & Fig.\ref{case: business} \\
\midrule
\multicolumn{11}{c}{\textit{Contradiction}} \\
\midrule
Negation & 200 & 4 & 1.0 & 0.0 & 12.8 & 1.0 & 0.00 & COCO & \makecell{Synthetic} & Fig.\ref{case: negation} \\
Vehicle Design & 88 & 700 & 1.0 & 152.5 & 107.5 & 1.0 & 0.04 & \makecell{\designqa} & \makecell{Design Rules} & Fig.\ref{case: design} \\
Traffic Case & 80 & 796 & 1.8 & 19.5 & 123.3 & 1.0 & 0.58 & \makecell{Synthetic} & \makecell{Driving\\Handbook} & Fig.\ref{case: traffic} \\
\bottomrule
\end{tabular}
\end{adjustbox}
\end{table}

\subsection{Task Formulation}


We define the task of multimodal retrieval as follows. Let $Q = \{q_1,\dots,q_n\}$ be the set of queries and $D = \{d_1,\dots,d_m\}$ the document corpus. Each query $q$ and document $d$ is represented as a sequence of segments $(x_1,\dots,x_k)$, where each segment $x$ can be either text or an image. For a query $q$, a document can be either a positive document $d_+$ (relevant) or a negative document $d_-$ (non-relevant). In reasoning-intensive retrieval, a document $d$ is considered relevant if it provides principles or theorems that support the reasoning chain required to answer query $q$ \cite{bright}. Unlike prior studies \cite{mieb,mmdocir}, we do not constrain the corpus to uniform data types, reflecting more realistic retrieval scenarios. To evaluate diverse reasoning capabilities, we design three types of retrieval tasks in \ours: 

\begin{itemize}[leftmargin=*]
    \item \textbf{\knowledge.} It emphasizes reasoning over broad expert domain knowledge. For a multimodal query, a document is relevant if expert annotators confirm that it contributes to reasoning about the query by providing critical concepts or theoretical foundations.
    
    \item \textbf{\theorem.} It targets the theorem-based reasoning over calculation problems. For a multimodal calculation query, a document is  relevant if it conveys the same underlying theorem or formula needed to solve the problem.
    
    \item \textbf{\contradiction.} It requires logical reasoning to detect conflicting or inconsistent concepts. For a multimodal case description query, a document is relevant if it provides the rule or requirement that the query violates.
\end{itemize}

\subsection{Knowledge: Retrieving web pages that help answer questions}
\label{construct knowledge}

\begin{figure}[t!]
    \centering
    \includegraphics[width=1\textwidth]{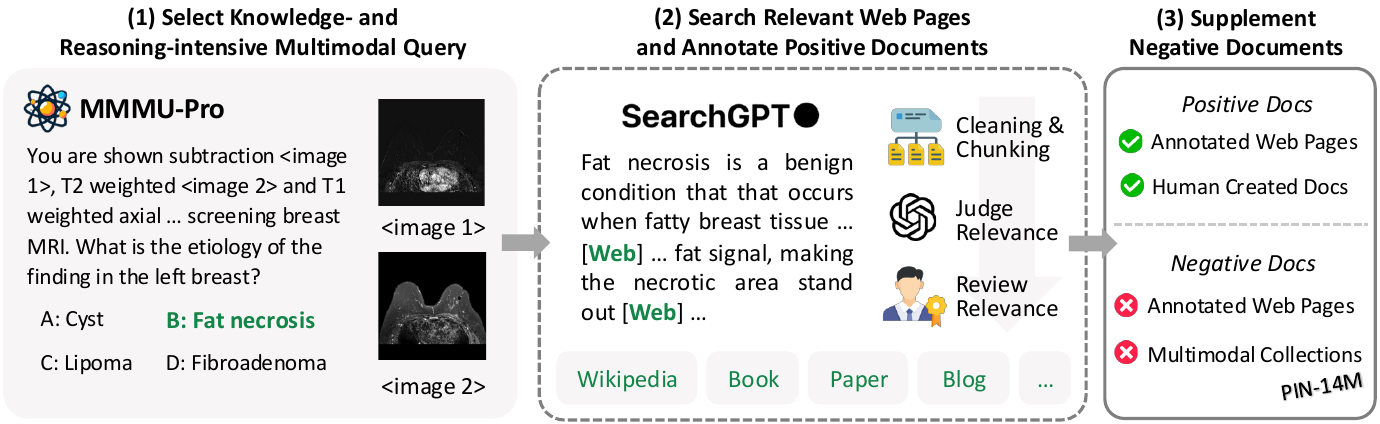}
    \captionsetup{justification=justified, singlelinecheck=false}  
    \caption{An overview of the data construction workflow for \ours (\knowledge). We select and convert knowledge- and reasoning-intensive questions from \mmmupro \cite{mmmupro} into retrieval queries. Web pages such as Wikipedia, blogs, and papers referenced by the \gptsearch model during reasoning are processed into documents through screen capturing, OCR \cite{monkeyocr}, and chunking. The relevance of resulting documents is first evaluated by GPT and then verified by expert annotators. Lastly, we source negative documents from the knowledge-intensive multimodal collection \pin \cite{pin} to construct a sizable corpus.
    }
    \label{pipeline}
\end{figure}

\mmmu \cite{mmmu} is one of the most widely used benchmark for evaluating multi-discipline multimodal understanding in MLLMs. Its robust version, \mmmupro \cite{mmmupro}, excludes questions solvable by text-only models, expands the candidate options, and provides verified correct answers. We repurpose the knowledge- and reasoning-intensive questions in \mmmupro as queries $Q$ and construct a corpus $D$ of image–text interleaved documents. The positive documents $D_+$ are scraped from relevant websites referenced by the \gptsearch\footnote{\gptsearch refers to the version \texttt{gpt-4o-search-preview-2025-03-11} throughout this work.} model \cite{gptsearch} and verified by human experts; while negative documents $D_-$ are augmented by sampling from the multimodal collection \pin \cite{pin} (see \autoref{pipeline}).

\paragraph{Selecting questions.} We prompt GPT-5\footnote{GPT-5 refers to the version \texttt{gpt-5-2025-08-07} throughout this work.} to categorize \mmmupro questions into two groups, \ie knowledge-based and calculation questions. We adopt calculation questions for the \theorem subset in Section~\ref{sec:theorm}. For knowledge questions, we then instruct GPT-5 to filter out questions that require only superficial reasoning over text and images, without reliance on external domain expertise. For the remaining questions, we generate detailed descriptions for each associated image using GPT-5, which we include as part of the input context for subsequent steps.


\paragraph{Constructing positive and hard negative documents.} 
Unlike keyword- or semantic-based multimodal retrieval benchmarks, collecting positive documents for our queries is more time-consuming because it requires identifying and validating multimodal sources that support the query’s answer. 
To address this, we design a semi-automated pipeline with human expert annotators.
Specifically, for each query, given the GPT-5-generated image descriptions and ground-truth answer, we prompt GPT-Search to reason over the question and generate an explanation for the correct answer with reference web links pointing to diverse materials such as Wikipedia, books, academic papers, and blogs. To preserve the completeness of multimodal content, we capture these webpages as PDFs, apply MonkeyOCR \cite{monkeyocr} to extract interleaved text and images, and split the content into chunks while preserving image references. Resulting documents are then screened by GPT-5 and validated by human experts about whether they support the correct answer. Documents with GPT-human agreement on relevance are retained as positives, those agreed irrelevant as hard negatives, while ambiguous cases (30–60\% across domains) are discarded. In cases where \gptsearch fails to retrieve relevant documents (38.2\% of questions), expert annotators are instructed to search the web and create one supporting document, optionally including image links within the text. Due to the complexity of the questions, the number of positive documents per query is typically fewer than four. We annotate data anonymously through the Turkle platform \cite{turkle2025}, with detailed guidelines provided in Appendix~\ref{appendix knowledge}.


\paragraph{Constructing additional negative documents.} After the previous step, we obtain 993 cleaned and annotated documents for 555 queries. To construct a sizable retrieval corpus comparable to \cite{mieb,bright}, we supplement these with negative documents sampled from the large-scale multimodal collection PIN-14M \cite{pin}, which contains knowledge-intensive resources such as medical articles from PubMed Central (PMC)\footnote{\url{https://www.ncbi.nlm.nih.gov/pmc/}}
 and web content from OBELICS \cite{obelics}. Given the wide topic coverage and large number of documents in PIN-14M, we assume a low probability of false negatives for our sampled documents. We validate this assumption through manual error analysis in Section~\ref{error}. In total, we curate a corpus of 26,223 documents, including text only, image only, and text-image interleaved.\footnote{ The corpus could be further expanded by sampling additional expert-domain documents, which naturally increases retrieval difficulty and the probability of false negatives. We leave it as future work.}

\subsection{Theorem: Retrieving relevant theorems that solve problems}
\label{sec:theorm}

As introduced by \citet{bright}, retrieving relevant theorem statements can assist users in solving new math or physics problems. We extend this formulation to the multimodal domain by leveraging challenging calculation problems from \mmmupro. In this setting, the query $q$ is a image-centric calculation problem, and the corpus $D$ consists of theorem descriptions across domains such as mathematics, physics, engineering, and business. A document $d$ is regarded as positive if it describes a theorem applicable to solving the query problem.


\paragraph{Selecting questions.} From the calculation questions in \mmmupro, we first use GPT-5 to exclude questions that explicitly state the required theorem in the  text. The remaining questions are then organized into four major domains: Math, Physics, Engineering, and Business. The Engineering domain further includes areas such as Mechanical Engineering, Computer Science, and Electronics, while the Business domain covers Finance, Economics, Marketing, and related fields. Then, we prompt GPT-5 to reason through each multimodal question, produce a final answer, and summarize the key theorems used in the solution. We exclude questions for which GPT-5 produces incorrect answers, with a final set of 512 questions.

\paragraph{Constructing positive and negative documents.} We adopt the theorem statements from \bright \cite{bright} as the primary retrieval corpus ($\sim$13.8k documents), reflecting the realistic setting where most theorems are expressed in text. For each question, the summarized key theorems are used as queries to retrieve the top-10 candidate statements from the corpus with the Qwen3-Embedding model~\cite{qwen3embedding}. Among these candidates, GPT-5 identifies the most relevant theorem statements, which are retained as positive documents, while the rest serve as negatives.

\paragraph{Constructing additional positive documents.} Not all theorems have relevant counterparts in \bright. To address this, we scrape additional theorem statements, optionally accompanied by illustrative images, from webpages such as Wikipedia, following the OCR pipeline described in Section~\ref{construct knowledge}. GPT-5 then rewrites these theorems to match the format of \bright statements. Finally, we deduplicate the scraped documents to ensure a consistent and complete retrieval corpus. Consequently, 63.6\% of the positive documents are sourced from webpages, with the remainder drawn from the \bright corpus. More details are presented in Appendix~\ref{appendix theorem}.

\subsection{Contradiction: Retrieving Contradictory Rules and Requirements}


Most existing datasets emphasize retrieving positively supporting evidence for a query \cite{mieb, infoseek, mmdocir}. However, retrieving contradictory information could be of great importance especially in expert domains. For example, a user may provide a case description and seek evidence of violation of laws, policies, or guidelines, as shown in \autoref{case: traffic}. In this setting, the query $q$ is a case description (\eg traffic or vehicle design cases), while the corpus $D$ comprises mandated rules (\eg driving theory handbooks or design requirements). A document $d$ is considered positive if it contains the statement or rule contradicting the query case. Unlike traditional retrieval tasks, this new formulation requires not only semantic matching between query and document but also deep logical reasoning to identify conflicting concepts.

\paragraph{Negation.} To study contradiction retrieval, we first design a synthetic task inspired by the negation benchmark NegBench \cite{vlm_not_undersatnd_negation}. Given an image from COCO \cite{coco} with ground truth object annotations, we synthesize four candidate text descriptions: three accurately reflecting the objects and one containing a contradiction, either by asserting the existence of a non-existent object or the absence of an existent one. The models are required to pinpoint the text description conflicting with the given image in a multi-choice setup. For example, in \autoref{case: negation}, the query image shows a keyboard on the table, while the positive document explicitly states that none is present, revealing a contradiction. More details are provided in Appendix~\ref{appendix negation}. 

\paragraph{Vehicle Design.} To evaluate contradiction retrieval in engineering documents, we construct a vehicle design task by leveraging the Formula SAE Rulebook and design cases from the DesignQA dataset \cite{designqa}. In industrial product design, designers must review hundreds of pages of requirement documents to ensure their designs comply with specifications. To assist designers, retrieval systems are expected to identify the specific sections that a design case fails to satisfy. For example, in \autoref{case: design}, the vehicle’s wheelbase in the design is shorter than the required minimum, indicating a contradiction. During data preparation, we introduce variations to the design cases and chunk the lengthy design document, as detailed in Appendix~\ref{appendix design}.

\paragraph{Traffic Case.} Retrieval systems have been applied to legal documents to assist legal professionals in preparing arguments and citations \cite{legal}. To evaluate this capability in multi-modality, we construct a traffic case task to assess whether retrievers can identify which driving rules are violated in traffic cases. We build the corpus by chunking official driving handbooks \cite{sg_driving} into sections. Meanwhile, we build the query set by selecting dozens of driving rules, each linked to several annotated violation cases. We augment these violation cases by replacing key textual elements with AI-generated images using Qwen-Image \cite{qwenimage}. For example, as shown in \autoref{case: traffic}, a car is driving only 3 meters behind the vehicle ahead — significantly less than the required safe distance. Further details are provided in Appendix~\ref{appendix traffic}.

\begin{table}[htbp]
\centering
\caption{The performance of retrieval models on \ours. We report nDCG@10 for all subtasks except Negation, for which we use Hit@1: Art, Medicine (Med.), Science (Sci.), Humanities (Hum.), Math, Physics (Phy.), Engineering (Eng.), Business (Bus.), Negation (Neg.), Design, and Traffic. Avg. denotes the average score across 11 subtasks. The best score on each subtask is highlighted in \textbf{bold}, and the second best is \underline{underlined}.}
\label{tab:main_results}
\begin{adjustbox}{width=\textwidth,center}
\begin{tabular}{lcccccccccccc}
\toprule
 & \multicolumn{4}{c}{Knowledge} & \multicolumn{4}{c}{Theorem} & \multicolumn{3}{c}{Contradiction} & Avg. \\
\cmidrule(lr){2-5} \cmidrule(lr){6-9} \cmidrule(lr){10-12}
Model & Art & Med. & Sci. & Hum.  & Math & Phy. & Eng. & Bus. & Neg. & Design & Traffic & \\
\midrule
\multicolumn{13}{c}{\textit{Text Models with Image Caption} (T2T)} \\
\midrule
BGE-M3 & 48.6 & 30.0 & 42.4 & 45.6 & 16.5 & 19.5 & 21.6 & 39.3 & 16.0 & 25.9 & 17.4 & 29.3\\
NV-Embed-v2 & 70.7 & 46.8 & 65.7 & 66.6 & 26.4 & 35.2 & \underline{32.9} & 52.2 & 12.5 & 42.1 & 42.2 & 44.8 \\
Qwen3-Embedding & \underline{71.9} & \textbf{53.2} & \textbf{72.5} & \textbf{74.4} & \textbf{37.7} & \textbf{50.2} & \textbf{42.9} & \textbf{58.3} & 12.0 & \textbf{67.8} & \textbf{54.2} & \textbf{54.1}\\
\midrule
\multicolumn{13}{c}{\textit{Text and Image Two-Stream Models with Vector Fusion} (IT2IT)} \\
\midrule
EVA-CLIP  & 10.2 & 13.5 & 26.1 & 12.9 & 6.2 & 12.2 & 10.7 & 17.4 & 8.5 & 4.4 & 5.4 & 11.6 \\
SigLIP & 26.7 & 14.7 & 26.7 & 12.3 & 7.4 & 6.5 & 5.9 & 12.5 & 13.5 & 4.9 & 9.6 & 12.8 \\
OpenCLIP & 56.0 & 17.9 & 33.2 & 22.0 & 7.5 & 6.6 & 7.3 & 14.0 & 13.0 & 8.1 & 12.4 & 18.0 \\
JinaCLIP & 21.4 & 16.8 & 27.1 & 10.7 & 10.9 & 7.5 & 9.1 & 13.7 & 10.5 & 16.5 & 9.7 & 14.0 \\
\midrule
\multicolumn{13}{c}{\textit{Multimodal Models with Merged Image} (IT2IT)} \\
\midrule
VISTA & 21.3 & 27.8 & 32.6 & 17.0 & 18.8 & 17.1 & 17.3 & 28.6 & \underline{20.0} & 20.2 & 9.4 & 20.9 \\
E5-V & 25.1 & 11.7 & 16.6 & 10.8 & 2.1 & 3.4 & 2.5 & 5.2 & 11.5 & 3.7 & 2.1 & 8.6 \\
MM-Embed & 65.6 & \underline{53.0} & 63.5 & 62.8 & 23.6 & 30.8 & 27.4 & 44.9 & 7.0 & 23.8 & 34.9 & 39.8 \\
VLM2Vec & 53.5 & 22.4 & 36.7 & 24.0 & 2.1 & 2.8 & 2.8 & 2.9 & 11.5 & 5.6 & 18.3 & 18.1 \\
GME-Qwen2-VL & 54.3 & 40.1 & 46.8 & 45.6 & \underline{28.8} & 36.0 & 30.2 & 45.1 & 15.0 & 26.3 & 29.6 & 36.2
\\
Ops-MM-Embedding &  \textbf{79.3}&52.5&\underline{70.0}& \underline{67.8}& 27.7 &  \underline{39.5} & 30.1 & \underline{52.3} & 8.0 & 55.9 &45.8& \underline{48.1} \\
\midrule
\multicolumn{13}{c}{\textit{Multimodal Models with Document as Image} (T2I)} \\
\midrule
GME-Qwen2-VL & 54.0 & 40.7 & 59.0 & 50.1 & 21.2 & 22.1 & 27.0 & 45.3 & 14.5 & 56.1& 40.1 & 39.1\\
Ops-MM-Embedding & 67.7 & 48.8 & 67.7 & 63.9 & 25.0 & 34.0 & 29.2 & 49.0 & 10.5 & \underline{59.8} & \underline{46.3} & 45.6 \\
ColPali  & 36.1 & 29.9 & 42.7 & 29.2& 7.3 & 17.5 & 13.5 & 34.6 & \textbf{28.5} & 19.4 & 18.2 & 25.2 \\

\bottomrule
\end{tabular}
\end{adjustbox}
\end{table}

\section{Experiments}
\label{sec: experiments}

\subsection{Experimental setup}

We evaluate 4 types of multimodal retrieval setups with 14 frontier models, as follows:
(1) \textbf{Text models with image caption (T2T)}: We assess text retrievers, namely BGE-M3 \cite{bge_m3}, NV-Embed-V2 \cite{nv_embed}, and Qwen3-Embedding-8B \cite{qwen3embedding}, by pairing with MLLM-generated image captions (see Appendix~\ref{experiment} for details). 
(2) \textbf{Text and image two-stream models with vector fusion (IT2IT)}: We evaluate CLIP-style two-stream models, including EVA-CLIP \cite{evaclip}, SigLIP \cite{siglip}, OpenCLIP \cite{openclip}, and JinaCLIP \cite{jinaclip}, by a simple vector-fusion strategy. Given an input sequence, we concatenate all text chunks for one text embedding $t$, while all images are concatenated vertically for another image embedding $i$. Following \mteb \cite{mieb}, the final score is computed using the fused embedding $e = t + i$. 
(3) \textbf{Multimodal models with merged image (IT2IT)}: We evaluate multimodal retrievers including VISTA \cite{vista}, E5-V \cite{e5v}, MM-Embed \cite{mmembed}, VLM2Vec \cite{vlm2vec}, Ops-MM-Embedding \cite{ops} and GME-Qwen2-VL \cite{gme}. Since these models support only single-image input, multiple images are concatenated in the same way as for two-stream models. 
(4) \textbf{Multimodal models with document as image (T2I)}: We also include the document retrieval paradigm that receives text-only query and encode entire multimodal documents as screenshot images, such as ColPali \cite{colpali}. Because these models are trained for text queries, query images are replaced with LLM-generated captions, similar to the text retriever setup. Besides, we note that a native image–text interleaved model, \tiir \cite{tiir}, has been introduced and is expected to best fit the interleaved format of \ours; however, it is not publicly available.
We provide details of each model in Appendix ~\ref{experiment}. Following prior work \cite{mieb, bright}, we use nDCG@10 as the main evaluation metric except Negation. Since each query in Negation has exactly one gold document among four candidates, we adopt Hit@1 as the main metric for this task.



\subsection{Main results}

\paragraph{Multimodal retrieval systems lag behind text retrieval-based approaches on knowledge- and reasoning-intensive images.}
As shown in \autoref{tab:main_results}, the text retriever Qwen3-Embedding combined with LLM-based image captioning achieves the highest performance (54.1 nDCG@10). Although captions may omit certain visual details, they provide rich contextual descriptions and additional knowledge that substantially benefit retrieval. In contrast, multimodal systems struggle with the expert-level query images in \ours, which often require deep reasoning, such as diagnosing microscopic tissue sections (\autoref{fig1}). CLIP-style two-stream models are particularly limited, as their training emphasizes alignment of superficial text–image semantics and model sizes are relatively small. The most recent MLLM-based embedding models, such as Ops-MM-Embedding, show promising results under both interleaved text–image and document-as-image paradigms, indicating the effectiveness of unified training on diverse retrieval tasks. 

\paragraph{Multimodal retrieval systems perform particularly poorly on reasoning-intensive tasks.}
While Ops-MM-Embedding achieves a solid 67.4 nDCG@10 on \knowledge subtasks, its performance drops sharply to 37.4 and 36.6 on \theorem and \contradiction, respectively. Models such as E5-V and VLM2Vec perform even worse, essentially failing on these tasks. This gap highlights the difficulty of extracting abstract concepts from practical problems, for example linking an image-based physics question to the relevant theorem in \autoref{fig1}. Notably, Hit@1 scores for most models on the synthetic \contradiction task Negation remain below 25\%—equivalent to random guessing given four candidates per query. As illustrated in the Negation example \autoref{case: negation}, humans can readily detect conflicting concepts embedded within supporting evidence, yet retrieval models struggle even for strong text embedding models. Although the candidate corpora for the Design and Traffic subtasks are much smaller than those of standard knowledge bases \cite{bright,mmdocir}, models still struggle to identify the underlying contradictions. Nevertheless, surface-level semantic matching remains useful in these settings, as it allows models to locate relevant documents without fully resolving the conflicting concepts (\eg a query concerning driving speed matched with a document specifying the speed limit). These findings suggest that current retrieval models possess strong capabilities in semantic matching and information seeking, but remain fundamentally limited in their reasoning ability.

\paragraph{Substantial differences in performance are evident across models and domains.} Across all four multimodal retrieval settings, we observe a wide performance difference between models. For instance, among multimodal models with merged image, the weakest model, E5-V, achieves only 8.6 nDCG@10, whereas Ops-MM-Embedding reaches 48.1 nDCG@10, revealing substantial methodological differences. As \ours is the first multidisciplinary multimodal retrieval benchmark, it enables fine-grained domain-level evaluation. For example, as shown in the breakdown performance \autoref{tab:breakdown_results}, MM-Embed performs competitively with Ops-MM-Embedding in medical domains such as Clinical Medicine and Diagnostics, yet lags behind in art-related tasks. We also observe pronounced variation in retrieval difficulty across domains. In the Art subtasks, systems can often succeed by matching query images to visually identical or similar artworks, which narrows the search space. However, in medical imaging, such overlap is rare, and models are required to identify underlying pathological and radiological features rather than relying on superficial visual similarity.

\section{Analysis}

\subsection{Qualitative analysis}
\label{error}

To understand model limitations, we conduct 30 case studies by manually reviewing their top-10 documents retrieved by Ops-MM-Embedding. There are two major failure patterns that we have observed. (1) \textbf{Visual bias over contextual relevance}: in the Agriculture case as shown in \autoref{case: fauna}, the model ranks a negative document higher because it contains a nematode SEM image resembling the earthworm image in the query, even though the positive document provides a detailed discussion of the key topic Fauna. Similar errors occur in Medicine, where visually similar eye images from different diseases mislead the model. (2) \textbf{Failure of higher-level deduction}: in the Traffic case as shown in \autoref{case: tunnel}, the model assigns a higher score to a negative document than to a positive one because both depict cars, tunnels, and lane markings. However, it fails to infer that the car is crossing the line, which contradicts the positive document’s instruction to ``Stay in lane''.
Although multimodal retrievers exhibit these shortcomings and lag behind text-only retrievers with image captions, we believe they remain essential because many real-world queries inherently span across modalities. Fundamentally, textual descriptions alone cannot fully capture the nuanced information in images, especially when MLLMs lack the required visual knowledge.

\subsection{Test-time scaling in retrieval}


Query expansion is a widely used technique, recently framed as test-time scaling in retrieval \cite{reasonir}. Prior work \cite{bright} demonstrates that incorporating explicit reasoning substantially improves performance on reasoning-intensive text retrieval tasks. Motivated by this, we have conducted comparative experiments to evaluate the effectiveness for multimodal retrieval. Specifically, we prompt MLLMs, including Qwen2-VL-2B-Instruct \cite{qwen2vl} and Qwen2.5-VL-72B-Instruct \cite{qwen2.5vl}, to generate reasoning traces, including question summarization and chain-of-thought reasoning, following \cite{bright}. As shown in \autoref{tab:test_scale}, replacing the original queries with MLLM-generated reasoning traces leads to substantial performance improvements: $+5.1$ for Qwen2-VL-2B and $+14.8$ for Qwen2.5-VL-72B. The improvements are particularly pronounced on \knowledge tasks, whereas \theorem tasks benefit to a lesser extent. Meanwhile, we observe that, without constraining output length, the larger model Qwen2.5-VL-72B produces on average 20\% and 66\% more tokens than Qwen2-VL-2B in \knowledge and \theorem respectively, trading higher inference cost for larger performance gains (see more details in Appendix~\ref{test-time-sacling-in-retrieval}).

\section{Conclusion}

We introduce \ours, a realistic, multidisciplinary, reasoning-intensive multimodal retrieval benchmark. We leverage knowledge- and reasoning-intensive questions from \mmmupro and build a sizable multimodal corpus with positive documents verified by human experts. In addition, we introduce Contradiction Retrieval for evaluating models' logical reasoning capabilities to identify conflicts. Comprehensive evaluation shows that multimodal retrieval systems lag behind their text-retrieval counterparts, indicating substantial room for improvement. Although state-of-the-art multimodal models excel in \knowledge domains, they drop nearly 30 points on reasoning-intensive tasks. We hope \ours facilitates identifying model limitations and advancing multimodal retrieval.

\section*{Code of Ethics and Ethics Statement}

All data used in constructing \ours are sourced from publicly available materials and are employed solely for academic research, not commercial use. We have carefully ensured that the dataset contains no private information or harmful content, such as discriminatory, violent, or unethical material. Our goal is to support socially beneficial research. Following the practice of MMMU \citep{mmmu}, our annotators and validators are instructed to avoid using materials from websites that prohibit copying or redistribution when reviewing \ours documents. Consequently, most documents are derived from sources that are free of copyright restrictions, such as Wikipedia pages, government reports (e.g., National Institutes of Health and Singapore Police Force), and PubMed Central (PMC). The datasets we build upon also carry permissive public licenses, including MMMU (Apache-2.0), PIN-14M (CC-BY-4.0), COCO (CC-BY-4.0), and BRIGHT (CC-BY-4.0). For test-time scaling, we primarily focus on text expansion rather than image resizing and process as the text expansion has shown more significant impacts.

\section*{Reproducibility}


Our datasets and annotation process are introduced in \autoref{sec: constructing ours}, and the experimental settings are described in \autoref{sec: experiments}. Specific implementation details are provided in Appendix~\ref{experiment}. To facilitate the reproduction of our experiments, the data are available at \url{https://huggingface.co/datasets/MRMRbenchmark}
, and the evaluation code is provided at \url{https://github.com/rebeccaz4/MRMR}
.

\section*{Acknowledgments}
This research is supported by the RIE2025 Industry Alignment Fund – Industry Collaboration Projects (IAF-ICP) (Award I2301E0026), administered by A*STAR, as well as supported by Alibaba Group and NTU Singapore through Alibaba-NTU Global e-Sustainability CorpLab (ANGEL). Siyue Zhang and Chen Zhao were supported by NYU Shanghai Center for Data Science. This work was supported in part through the NYU IT High Performance Computing resources, services, and staff expertise.

\bibliography{iclr2026_conference}

@misc{vidore,
      title={ViDoRe Benchmark V2: Raising the Bar for Visual Retrieval}, 
      author={Quentin Macé and António Loison and Manuel Faysse},
      year={2025},
      url={https://arxiv.org/abs/2505.17166}, 
}

@article{qwen3embedding,
  title={Qwen3 Embedding: Advancing Text Embedding and Reranking Through Foundation Models},
  author={Zhang, Yanzhao and Li, Mingxin and Long, Dingkun and Zhang, Xin and Lin, Huan and Yang, Baosong and Xie, Pengjun and Yang, An and Liu, Dayiheng and Lin, Junyang and Huang, Fei and Zhou, Jingren},
  journal={arXiv preprint arXiv:2506.05176},
    url={https://arxiv.org/abs/2506.05176},
  year={2025}
}

@misc{nv_embed,
      title={NV-Embed: Improved Techniques for Training LLMs as Generalist Embedding Models}, 
      author={Chankyu Lee and Rajarshi Roy and Mengyao Xu and Jonathan Raiman and Mohammad Shoeybi and Bryan Catanzaro and Wei Ping},
      year={2025},
      url={https://arxiv.org/abs/2405.17428}, 
}

@article{reasonir,
      title={ReasonIR: Training Retrievers for Reasoning Tasks}, 
      author={Rulin Shao and Rui Qiao and Varsha Kishore and Niklas Muennighoff and Xi Victoria Lin and Daniela Rus and Bryan Kian Hsiang Low and Sewon Min and Wen-tau Yih and Pang Wei Koh and Luke Zettlemoyer},
      journal={Proceedings of Conference on Language Modeling},
      year={2025},
      url={https://arxiv.org/abs/2504.20595}, 
}

@inproceedings{fashioniq,
      title={Fashion IQ: A New Dataset Towards Retrieving Images by Natural Language Feedback}, 
      author={Xin Zhang and Yanzhao Zhang and Wen Xie and Mingxin Li and Ziqi Dai and Dingkun Long and Pengjun Xie and Meishan Zhang and Wenjie Li and Min Zhang},
      booktitle={The IEEE / CVF Computer Vision and Pattern Recognition Conference (CVPR)},
      year={2021},
      url={https://openaccess.thecvf.com/content/CVPR2021/papers/Wu_Fashion_IQ_A_New_Dataset_Towards_Retrieving_Images_by_Natural_CVPR_2021_paper.pdf},
}

@InProceedings{cirr,
    author    = {Liu, Zheyuan and Rodriguez-Opazo, Cristian and Teney, Damien and Gould, Stephen},
    title     = {Image Retrieval on Real-Life Images With Pre-Trained Vision-and-Language Models},
    booktitle = {Proceedings of the IEEE/CVF International Conference on Computer Vision (ICCV)},
    year      = {2021},
    url = {https://openaccess.thecvf.com/content/ICCV2021/papers/Liu_Image_Retrieval_on_Real-Life_Images_With_Pre-Trained_Vision-and-Language_Models_ICCV_2021_paper.pdf}
}

@inproceedings{magiclens,
title={MagicLens: Self-Supervised Image Retrieval with Open-Ended Instructions},
author={Zhang, Kai and Luan, Yi and Hu, Hexiang and Lee, Kenton and Qiao, Siyuan and Chen, Wenhu and Su, Yu and Chang, Ming-Wei},
booktitle={The Forty-first International Conference on Machine Learning (ICML)},
year={2024},
url={https://arxiv.org/abs/2403.19651}
}

@misc{qwenimage,
      title={Qwen-Image Technical Report}, 
      author={Chenfei Wu and Jiahao Li and Jingren Zhou and Junyang Lin and Kaiyuan Gao and Kun Yan and Sheng-ming Yin and Shuai Bai and Xiao Xu and Yilei Chen and Yuxiang Chen and Zecheng Tang and Zekai Zhang and Zhengyi Wang and An Yang and Bowen Yu and Chen Cheng and Dayiheng Liu and Deqing Li and Hang Zhang and Hao Meng and Hu Wei and Jingyuan Ni and Kai Chen and Kuan Cao and Liang Peng and Lin Qu and Minggang Wu and Peng Wang and Shuting Yu and Tingkun Wen and Wensen Feng and Xiaoxiao Xu and Yi Wang and Yichang Zhang and Yongqiang Zhu and Yujia Wu and Yuxuan Cai and Zenan Liu},
      year={2025},
      url={https://arxiv.org/abs/2508.02324}, 
}

@misc{sg_driving,
  author       = {{Singapore Police Force}},
  title        = {Basic Theory of Driving},
  year         = {2017},
  note         = {Accessed: 2025-09-21},
  url          = {https://www.police.gov.sg/~/media/spf/files/tp/online%20learning%20portal/bt%20eng%209th%20edition%20130717.pdf}
}

@article{designqa,
  title={DesignQA: A multimodal benchmark for evaluating large language models’ understanding of engineering documentation},
  author={Doris, Anna C and Grandi, Daniele and Tomich, Ryan and Alam, Md Ferdous and Ataei, Mohammadmehdi and Cheong, Hyunmin and Ahmed, Faez},
  journal={Journal of Computing and Information Science in Engineering},
  year={2025},
  publisher={American Society of Mechanical Engineers}
}

@misc{rader,
      title={RaDeR: Reasoning-aware Dense Retrieval Models}, 
      author={Debrup Das and Sam O' Nuallain and Razieh Rahimi},
      year={2025},
      url={https://arxiv.org/abs/2505.18405}, 
}

@misc{bge_reasoner,
  title        = {BGE-Reasoner: Towards End-to-End Reasoning-Intensive Information Retrieval},
  author       = {{FlagEmbedding}},
  year         = {2025},
  howpublished = {\url{https://github.com/FlagOpen/FlagEmbedding/tree/master/research/BGE_Reasoner}},
  note         = {Accessed: 2025-09-12}
}

@inproceedings{legal,
    title = "Legal Case Retrieval: A Survey of the State of the Art",
    author = "Feng, Yi  and
      Li, Chuanyi  and
      Ng, Vincent",
    booktitle = "Proceedings of the 62nd Annual Meeting of the Association for Computational Linguistics (Volume 1: Long Papers)",
    year = "2024",
    url = "https://aclanthology.org/2024.acl-long.350/",
}

@article{mragbench,
          title={MRAG-Bench: Vision-Centric Evaluation for Retrieval-Augmented Multimodal Models},
          author={Hu, Wenbo and Gu, Jia-Chen and Dou, Zi-Yi and Fayyaz, Mohsen and Lu, Pan and Chang, Kai-Wei and Peng, Nanyun},
          journal={Proceedings of The International Conference on Learning Representations (ICLR)},
        url={https://openreview.net/forum?id=Usklli4gMc},
          year={2025}
}

@article{vlm2vec,
          title={VLM2Vec: Training Vision-Language Models for Massive Multimodal Embedding Tasks},
           author={Ziyan Jiang and Rui Meng and Xinyi Yang and Semih Yavuz and Yingbo Zhou and Wenhu Chen},
          journal={Proceedings of The International Conference on Learning Representations (ICLR)},
        url={https://openreview.net/forum?id=TE0KOzWYAF},
          year={2025}
}

@misc{mieb,
      title={MIEB: Massive Image Embedding Benchmark}, 
      author={Chenghao Xiao and Isaac Chung and Imene Kerboua and Jamie Stirling and Xin Zhang and Márton Kardos and Roman Solomatin and Noura Al Moubayed and Kenneth Enevoldsen and Niklas Muennighoff},
      year={2025},
      url={https://arxiv.org/abs/2504.10471}, 
}

@inproceedings{scimmir,
    title = "{S}ci{MMIR}: Benchmarking Scientific Multi-modal Information Retrieval",
    author = "Wu, Siwei  and
      Li, Yizhi  and
      Zhu, Kang  and
      Zhang, Ge  and
      Liang, Yiming  and
      Ma, Kaijing  and
      Xiao, Chenghao  and
      Zhang, Haoran  and
      Yang, Bohao  and
      Chen, Wenhu  and
      Huang, Wenhao  and
      Al Moubayed, Noura  and
      Fu, Jie  and
      Lin, Chenghua",
    booktitle = "Findings of the Association for Computational Linguistics: ACL 2024",
    year = "2024",
    url = "https://aclanthology.org/2024.findings-acl.746/",
}

@misc{rank1,
      title={Rank1: Test-Time Compute for Reranking in Information Retrieval}, 
      author={Orion Weller and Kathryn Ricci and Eugene Yang and Andrew Yates and Dawn Lawrie and Benjamin Van Durme},
      year={2025},
      journal={Proceedings of Conference on Language Modeling},
      url={https://arxiv.org/abs/2502.18418}, 
}

@misc{diver,
      title={DIVER: A Multi-Stage Approach for Reasoning-intensive Information Retrieval}, 
      author={Meixiu Long and Duolin Sun and Dan Yang and Junjie Wang and Yue Shen and Jian Wang and Peng Wei and Jinjie Gu and Jiahai Wang},
      year={2025},
      url={https://arxiv.org/abs/2508.07995}, 
}

@inproceedings{gme,
      title={GME: Improving Universal Multimodal Retrieval by Multimodal LLMs}, 
      author={Xin Zhang and Yanzhao Zhang and Wen Xie and Mingxin Li and Ziqi Dai and Dingkun Long and Pengjun Xie and Meishan Zhang and Wenjie Li and Min Zhang},
      booktitle={The IEEE / CVF Computer Vision and Pattern Recognition Conference (CVPR)},
      year={2025},
      url={https://openaccess.thecvf.com/content/CVPR2025/papers/Zhang_Bridging_Modalities_Improving_Universal_Multimodal_Retrieval_by_Multimodal_Large_Language_CVPR_2025_paper.pdf},
}

@inproceedings{uniir,
  title={Uniir: Training and benchmarking universal multimodal information retrievers},
  author={Wei, Cong and Chen, Yang and Chen, Haonan and Hu, Hexiang and Zhang, Ge and Fu, Jie and Ritter, Alan and Chen, Wenhu},
    booktitle = "The European Conference on Computer Vision (ECCV)",
    year = "2024",
    url = "https://www.ecva.net/papers/eccv_2024/papers_ECCV/papers/11927.pdf"
}

@inproceedings{tiir,
    title = "Towards Text-Image Interleaved Retrieval",
    author = "Zhang, Xin  and
      Dai, Ziqi  and
      Li, Yongqi  and
      Zhang, Yanzhao  and
      Long, Dingkun  and
      Xie, Pengjun  and
      Zhang, Meishan  and
      Yu, Jun  and
      Li, Wenjie  and
      Zhang, Min",
    booktitle = "Proceedings of the 63rd Annual Meeting of the Association for Computational Linguistics (Volume 1: Long Papers)",
    year = "2025",
    url = "https://aclanthology.org/2025.acl-long.214/"
}

@inproceedings{mmmu,
title={MMMU: A Massive Multi-discipline Multimodal Understanding and Reasoning Benchmark for Expert AGI},
author={Xiang Yue and Yuansheng Ni and Kai Zhang and Tianyu Zheng and Ruoqi Liu and Ge Zhang and Samuel Stevens and Dongfu Jiang and Weiming Ren and Yuxuan Sun and Cong Wei and Botao Yu and Ruibin Yuan and Renliang Sun and Ming Yin and Boyuan Zheng and Zhenzhu Yang and Yibo Liu and Wenhao Huang and Huan Sun and Yu Su and Wenhu Chen},
booktitle={Proceedings of CVPR},
year={2024},
}

@inproceedings{infoseek,
    title = "Can Pre-trained Vision and Language Models Answer Visual Information-Seeking Questions?",
    author = "Chen, Yang  and
      Hu, Hexiang  and
      Luan, Yi  and
      Sun, Haitian  and
      Changpinyo, Soravit  and
      Ritter, Alan  and
      Chang, Ming-Wei",
    booktitle = "Proceedings of the 2023 Conference on Empirical Methods in Natural Language Processing",
    year = "2023",
    url = "https://aclanthology.org/2023.emnlp-main.925/",

}

@inproceedings{unifying,
    title = "Unifying Multimodal Retrieval via Document Screenshot Embedding",
    author = "Ma, Xueguang  and
      Lin, Sheng-Chieh  and
      Li, Minghan  and
      Chen, Wenhu  and
      Lin, Jimmy",
    booktitle = "Proceedings of the 2024 Conference on Empirical Methods in Natural Language Processing",
    year = "2024",
    url = "https://aclanthology.org/2024.emnlp-main.373/",
}

@inproceedings{oven,
  title={Open-domain Visual Entity Recognition: Towards Recognizing Millions of Wikipedia Entities},
    author={Hu, Hexiang and Luan, Yi and Chen, Yang and Khandelwal, Urvashi and Joshi, Mandar and Lee, Kenton and Toutanova, Kristina and Chang, Ming-Wei},
    booktitle = {Proceedings of the IEEE/CVF International Conference on Computer Vision (ICCV)},
  year={2023},
url={https://arxiv.org/abs/2302.11154}
}

@misc{mmdocir,
      title={MMDocIR: Benchmarking Multi-Modal Retrieval for Long Documents}, 
      author={Kuicai Dong and Yujing Chang and Xin Deik Goh and Dexun Li and Ruiming Tang and Yong Liu},
      year={2025},
      url={https://arxiv.org/abs/2501.08828}, 
}

@article{gpt2,
  title={Language Models are Unsupervised Multitask Learners},
  author={Radford, Alec and Wu, Jeff and Child, Rewon and Luan, David and Amodei, Dario and Sutskever, Ilya},
  year={2019},
url={https://d4mucfpksywv.cloudfront.net/better-language-models/language-models.pdf}
}

@inproceedings{mmmupro,
    title = "{MMMU}-Pro: A More Robust Multi-discipline Multimodal Understanding Benchmark",
    author = "Yue, Xiang  and
      Zheng, Tianyu  and
      Ni, Yuansheng  and
      Wang, Yubo  and
      Zhang, Kai  and
      Tong, Shengbang  and
      Sun, Yuxuan  and
      Yu, Botao  and
      Zhang, Ge  and
      Sun, Huan  and
      Su, Yu  and
      Chen, Wenhu  and
      Neubig, Graham",
    booktitle = "Proceedings of the 63rd Annual Meeting of the Association for Computational Linguistics (Volume 1: Long Papers)",
    year = "2025",
    url = "https://aclanthology.org/2025.acl-long.736/",
}

@misc{diffembed,
      title={Diffusion vs. Autoregressive Language Models: A Text Embedding Perspective}, 
      author={Siyue Zhang and Yilun Zhao and Liyuan Geng and Arman Cohan and Anh Tuan Luu and Chen Zhao},
      year={2025},
      url={https://arxiv.org/abs/2505.15045}, 
}

@inproceedings{bright,
  title={BRIGHT: A Realistic and Challenging Benchmark for Reasoning-Intensive Retrieval},
  author={Su, Hongjin and Yen, Howard and Xia, Mengzhou and Shi, Weijia and Muennighoff, Niklas and Wang, Han-yu and Liu, Haisu and Shi, Quan and Siegel, Zachary S. and Tang, Michael and Sun, Ruoxi and Yoon, Jinsung and Arik, Sercan O. and Chen, Danqi and Yu, Tao},
  booktitle={International Conference on Learning Representations (ICLR)},
  year={2025},
  url={https://openreview.net/forum?id=ykuc5q381b}
}

@misc{evaclip,
      title={EVA-CLIP: Improved Training Techniques for CLIP at Scale}, 
      author={Quan Sun and Yuxin Fang and Ledell Wu and Xinlong Wang and Yue Cao},
      year={2023},
      url={https://arxiv.org/abs/2303.15389}, 
}

@misc{qwen2vl,
      title={Qwen2-VL: Enhancing Vision-Language Model's Perception of the World at Any Resolution}, 
      author={Peng Wang and Shuai Bai and Sinan Tan and Shijie Wang and Zhihao Fan and Jinze Bai and Keqin Chen and Xuejing Liu and Jialin Wang and Wenbin Ge and Yang Fan and Kai Dang and Mengfei Du and Xuancheng Ren and Rui Men and Dayiheng Liu and Chang Zhou and Jingren Zhou and Junyang Lin},
      year={2024},
      url={https://arxiv.org/abs/2409.12191}, 
}

@misc{qwen2.5vl,
      title={Qwen2.5-VL Technical Report}, 
      author={Shuai Bai and Keqin Chen and Xuejing Liu and Jialin Wang and Wenbin Ge and Sibo Song and Kai Dang and Peng Wang and Shijie Wang and Jun Tang and Humen Zhong and Yuanzhi Zhu and Mingkun Yang and Zhaohai Li and Jianqiang Wan and Pengfei Wang and Wei Ding and Zheren Fu and Yiheng Xu and Jiabo Ye and Xi Zhang and Tianbao Xie and Zesen Cheng and Hang Zhang and Zhibo Yang and Haiyang Xu and Junyang Lin},
      year={2025},
      url={https://arxiv.org/abs/2502.13923}, 
}

@inproceedings{dao2022flashattention,
  title={Flash{A}ttention: Fast and Memory-Efficient Exact Attention with {IO}-Awareness},
  author={Dao, Tri and Fu, Daniel Y. and Ermon, Stefano and Rudra, Atri and R{\'e}, Christopher},
  booktitle={Advances in Neural Information Processing Systems (NeurIPS)},
  year={2022}
}

@inproceedings{mteb,
    title = "{MTEB}: Massive Text Embedding Benchmark",
    author = "Muennighoff, Niklas  and
      Tazi, Nouamane  and
      Magne, Loic  and
      Reimers, Nils",
    booktitle = "Proceedings of the 17th Conference of the European Chapter of the Association for Computational Linguistics",
    year = "2023",
    url = "https://aclanthology.org/2023.eacl-main.148/",

}

@misc{jinaclip,
      title={jina-clip-v2: Multilingual Multimodal Embeddings for Text and Images}, 
      author={Andreas Koukounas and Georgios Mastrapas and Bo Wang and Mohammad Kalim Akram and Sedigheh Eslami and Michael Günther and Isabelle Mohr and Saba Sturua and Scott Martens and Nan Wang and Han Xiao},
      year={2024},
      url={https://arxiv.org/abs/2412.08802}, 
}

@inproceedings{obelics,
title={{OBELICS}: An Open Web-Scale Filtered Dataset of Interleaved Image-Text Documents},
author={Hugo Lauren{\c{c}}on and Lucile Saulnier and Leo Tronchon and Stas Bekman and Amanpreet Singh and Anton Lozhkov and Thomas Wang and Siddharth Karamcheti and Alexander M Rush and Douwe Kiela and Matthieu Cord and Victor Sanh},
booktitle={Thirty-seventh Conference on Neural Information Processing Systems Datasets and Benchmarks Track},
year={2023},
url={https://openreview.net/forum?id=SKN2hflBIZ}
}

@inproceedings{blip2,
author = {Li, Junnan and Li, Dongxu and Savarese, Silvio and Hoi, Steven},
title = {BLIP-2: bootstrapping language-image pre-training with frozen image encoders and large language models},
year = {2023},
booktitle = {Proceedings of the 40th International Conference on Machine Learning},
rul={https://dl.acm.org/doi/10.5555/3618408.3619222}
}

@inproceedings{siglip,
  title={Sigmoid loss for language image pre-training},
  author={Zhai, Xiaohua and Mustafa, Basil and Kolesnikov, Alexander and Beyer, Lucas},
  booktitle={Proceedings of the IEEE/CVF international conference on computer vision},
  year={2023},
url={https://arxiv.org/pdf/2303.15343}

}

@inproceedings{openclip,
  title={Reproducible scaling laws for contrastive language-image learning},
  author={Cherti, Mehdi and Beaumont, Romain and Wightman, Ross and Wortsman, Mitchell and Ilharco, Gabriel and Gordon, Cade and Schuhmann, Christoph and Schmidt, Ludwig and Jitsev, Jenia},
  booktitle={Proceedings of the IEEE/CVF Conference on Computer Vision and Pattern Recognition},
url={https://github.com/mlfoundations/open_clip},
  year={2023}
}

@inproceedings{clip,
  title={Learning Transferable Visual Models From Natural Language Supervision},
  author={Radford, Alec and Kim, Jong Wook and Hallacy, Chris and Ramesh, Aditya and Goh, Gabriel and Agarwal, Sandhini and Sastry, Girish and Askell, Amanda and Mishkin, Pamila and Clark, Jack and Krueger, Gretchen and Sutskever, Ilya},
  booktitle={Proceedings of the 38th International Conference on Machine Learning (ICML)},
  year={2021},
url={https://proceedings.mlr.press/v139/radford21a/radford21a.pdf}
}

@misc{bge_m3,
      title={BGE M3-Embedding: Multi-Lingual, Multi-Functionality, Multi-Granularity Text Embeddings Through Self-Knowledge Distillation}, 
      author={Jianlv Chen and Shitao Xiao and Peitian Zhang and Kun Luo and Defu Lian and Zheng Liu},
      year={2024},
url={https://huggingface.co/BAAI/bge-m3}

}

@article{pin,
  author    = {Junjie Wang and
               Yuxiang Zhang and
               Minghao Liu and
               Yin Zhang and
               Yatai Ji and
               Weihao Xuan and
               Nie Lin and
               Kang Zhu and
               Zhiqiang Lin and
               Yiming Ren and
               Chunyang Jiang and
               Yiyao Yu and
               Zekun Wang and
               Tiezhen Wang and
               Wenhao Huang and
               Jie Fu and
               Qunshu Lin and
               Yujiu Yang and
               Ge Zhang and
               Ruibin Yuan and
               Bei Chen and
               Wenhu Chen},
  title     = {{PIN:} {A} Knowledge-Intensive Dataset for Paired and Interleaved Multimodal Documents},
  year      = {2024},
url={https://huggingface.co/datasets/m-a-p/PIN-14M}
}

@inproceedings{vista,
    title = "{VISTA}: Visualized Text Embedding For Universal Multi-Modal Retrieval",
    author = "Zhou, Junjie  and
      Liu, Zheng  and
      Xiao, Shitao  and
      Zhao, Bo  and
      Xiong, Yongping",
    booktitle = "Proceedings of the 62nd Annual Meeting of the Association for Computational Linguistics (Volume 1: Long Papers)",
    year = "2024",
    url = "https://aclanthology.org/2024.acl-long.175/",
}

@InProceedings{circo,
      title={Zero-Shot Composed Image Retrieval with Textual Inversion}, 
      author={Alberto Baldrati and Lorenzo Agnolucci and Marco Bertini and Alberto Del Bimbo},
      year={2023},
booktitle = {Proceedings of the IEEE/CVF International Conference on Computer Vision (ICCV)},
    url={https://openaccess.thecvf.com/content/ICCV2023/papers/Baldrati_Zero-Shot_Composed_Image_Retrieval_with_Textual_Inversion_ICCV_2023_paper.pdf}
}

@misc{ops,
  title        = {OpenSearch-AI/Ops-MM-embedding-v1-7B},
  author       = {{OpenSearch-AI}},
  year         = {2025},
  url         ={https://huggingface.co/OpenSearch-AI/Ops-MM-embedding-v1-7B},
}

@misc{monkeyocr,
      title={MonkeyOCR: Document Parsing with a Structure-Recognition-Relation Triplet Paradigm}, 
      author={Zhang Li and Yuliang Liu and Qiang Liu and Zhiyin Ma and Ziyang Zhang and Shuo Zhang and Zidun Guo and Jiarui Zhang and Xinyu Wang and Xiang Bai},
      year={2025},
      url={https://arxiv.org/abs/2506.05218}, 
}

@misc{gptsearch,
  author       = {OpenAI},
  title        = {Introducing ChatGPT Search},
  howpublished = {\url{https://openai.com/index/introducing-chatgpt-search/}},
  note         = {Accessed: 2025-09-17},
  year         = {2024},
}

@inproceedings{edis,
    title = "{EDIS}: Entity-Driven Image Search over Multimodal Web Content",
  author={Siqi Liu and Weixi Feng and Tsu-jui Fu and Wenhu Chen and William Yang Wang},
    booktitle = "Proceedings of the 2023 Conference on Empirical Methods in Natural Language Processing",
    year = "2023",
      url={https://arxiv.org/abs/2305.13631}, 
}

@misc{
e5v,
title={E5-V: Universal Embeddings with Multimodal Large Language Models},
author={Ting Jiang and Shaohan Huang and Minghui Song and Zihan Zhang and Haizhen Huang and Liang Wang and Furu Wei and Weiwei Deng and Feng Sun and Qi Zhang and deqing wang and Fuzhen Zhuang},
year={2025},
url={https://openreview.net/forum?id=rD6LQagatR}
}

@misc{realmmragrealworldmultimodalretrieval,
      title={REAL-MM-RAG: A Real-World Multi-Modal Retrieval Benchmark}, 
      author={Navve Wasserman and Roi Pony and Oshri Naparstek and Adi Raz Goldfarb and Eli Schwartz and Udi Barzelay and Leonid Karlinsky},
      year={2025},
      url={https://arxiv.org/abs/2502.12342}, 
}

@inproceedings{coco,
  title={Microsoft coco: Common objects in context},
  author={Lin, Tsung-Yi and Maire, Michael and Belongie, Serge and Hays, James and Perona, Pietro and Ramanan, Deva and Doll{\'a}r, Piotr and Zitnick, C Lawrence},
  booktitle={European conference on computer vision},
  year={2014},
  url={https://cocodataset.org/images/coco-paper.png}
}

@inproceedings{vlm_not_undersatnd_negation,
  title={Vision-Language Models Do Not Understand Negation},
  author={Alhamoud, Kumail and Alshammari, Shaden and Tian, Yonglong and Li, Guohao and Torr, Philip and Kim, Yoon and Ghassemi, Marzyeh},
      booktitle={Proceedings of the IEEE/CVF Conference on Computer Vision and Pattern Recognition (CVPR)},
      year={2025},
      url={https://openaccess.thecvf.com/content/CVPR2025/papers/Alhamoud_Vision-Language_Models_Do_Not_Understand_Negation_CVPR_2025_paper.pdf}
}

@inproceedings{colpali,
      title={ColPali: Efficient Document Retrieval with Vision Language Models}, 
      author={Manuel Faysse and Hugues Sibille and Tony Wu and Bilel Omrani and Gautier Viaud and Céline Hudelot and Pierre Colombo},
  booktitle={International Conference on Learning Representations (ICLR)},
  year={2025},
  url={https://openreview.net/forum?id=ogjBpZ8uSi}
}

@misc{mmrag,
      title={MMRAG: Multi-Mode Retrieval-Augmented Generation with Large Language Models for Biomedical In-Context Learning}, 
      author={Zaifu Zhan and Jun Wang and Shuang Zhou and Jiawen Deng and Rui Zhang},
      year={2025},
      url={https://arxiv.org/abs/2502.15954}, 
}

@inproceedings{mmembed,
  title={MM-Embed: Universal Multimodal Retrieval with Multimodal LLMs},
  author={Sheng-Chieh Lin and Chankyu Lee and Mohammad Shoeybi and Jimmy Lin and Bryan Catanzaro and Wei Ping},
  booktitle={International Conference on Learning Representations (ICLR)},
  year={2025},
  url={https://openreview.net/forum?id=i45NQb2iKO}
}

@inproceedings{mmsearch,
      title={MMSearch: Benchmarking the Potential of Large Models as Multi-modal Search Engines}, 
      author={Dongzhi Jiang and Renrui Zhang and Ziyu Guo and Yanmin Wu and Jiayi Lei and Pengshuo Qiu and Pan Lu and Zehui Chen and Chaoyou Fu and Guanglu Song and Peng Gao and Yu Liu and Chunyuan Li and Hongsheng Li},
      booktitle={International Conference on Learning Representations (ICLR)},
      year={2025},
      url={https://arxiv.org/abs/2409.12959}, 
}

@misc{mmsearch_r1,
      title={MMSearch-R1: Incentivizing LMMs to Search}, 
      author={Jinming Wu and Zihao Deng and Wei Li and Yiding Liu and Bo You and Bo Li and Zejun Ma and Ziwei Liu},
      year={2025},
      url={https://arxiv.org/abs/2506.20670}, 
}

@misc{chromadb,
  title        = {ChromaDB: An open-source vector embedding database},
  author       = {Chroma},
  year         = {2025},
  url          = {https://github.com/chroma-core/chroma},
  note         = {Apache 2.0 license}
}

@misc{websearcher,
      title={WebResearcher: Unleashing unbounded reasoning capability in Long-Horizon Agents}, 
      author={Zile Qiao and Guoxin Chen and Xuanzhong Chen and Donglei Yu and Wenbiao Yin and Xinyu Wang and Zhen Zhang and Baixuan Li and Huifeng Yin and Kuan Li and Rui Min and Minpeng Liao and Yong Jiang and Pengjun Xie and Fei Huang and Jingren Zhou},
      year={2025},
      url={https://arxiv.org/abs/2509.13309}, 
}

@misc{medrag,
      title={MedRAG: Enhancing Retrieval-augmented Generation with Knowledge Graph-Elicited Reasoning for Healthcare Copilot}, 
      author={Xuejiao Zhao and Siyan Liu and Su-Yin Yang and Chunyan Miao},
      year={2025},
      url={https://arxiv.org/abs/2502.04413}, 
}

@inproceedings{medagents,
    title = "{M}ed{A}gents: Large Language Models as Collaborators for Zero-shot Medical Reasoning",
    author = "Tang, Xiangru  and
      Zou, Anni  and
      Zhang, Zhuosheng  and
      Li, Ziming  and
      Zhao, Yilun  and
      Zhang, Xingyao  and
      Cohan, Arman  and
      Gerstein, Mark",
    booktitle = "Findings of the Association for Computational Linguistics: ACL 2024",
    year = "2024",
    url = "https://aclanthology.org/2024.findings-acl.33/",
}

@inproceedings{finrag,
    title = "{G}raph{RAG}: Leveraging Graph-Based Efficiency to Minimize Hallucinations in {LLM}-Driven {RAG} for Finance Data",
    author = "Barry, Mariam  and
      Caillaut, Gaetan  and
      Halftermeyer, Pierre  and
      Qader, Raheel  and
      Mouayad, Mehdi  and
      Le Deit, Fabrice  and
      Cariolaro, Dimitri  and
      Gesnouin, Joseph",
    booktitle = "Proceedings of the Workshop on Generative AI and Knowledge Graphs (GenAIK)",
    year = "2025",
    url = "https://aclanthology.org/2025.genaik-1.6/",

}

@misc{humanitysexam,
      title={Humanity's Last Exam}, 
      author={Long Phan and Alice Gatti and Ziwen Han and Nathaniel Li and Josephina Hu and Hugh Zhang and Chen Bo Calvin Zhang and others},
      year={2025},
      url={https://arxiv.org/abs/2501.14249}, 
}

@misc{doubao,
      title={Seedream 2.0: A Native Chinese-English Bilingual Image Generation Foundation Model}, 
      author={Lixue Gong and Xiaoxia Hou and Fanshi Li and Liang Li and Xiaochen Lian and Fei Liu and Liyang Liu and Wei Liu and Wei Lu and Yichun Shi and Shiqi Sun and Yu Tian and Zhi Tian and Peng Wang and Xun Wang and Ye Wang and Guofeng Wu and Jie Wu and Xin Xia and Xuefeng Xiao and Linjie Yang and Zhonghua Zhai and Xinyu Zhang and Qi Zhang and Yuwei Zhang and Shijia Zhao and Jianchao Yang and Weilin Huang},
      year={2025},
      url={https://arxiv.org/abs/2503.07703}, 
}

@misc{browsecomp-plus,
      title={BrowseComp-Plus: A More Fair and Transparent Evaluation Benchmark of Deep-Research Agent}, 
      author={Zijian Chen and Xueguang Ma and Shengyao Zhuang and Ping Nie and Kai Zou and Andrew Liu and Joshua Green and Kshama Patel and Ruoxi Meng and Mingyi Su and Sahel Sharifymoghaddam and Yanxi Li and Haoran Hong and Xinyu Shi and Xuye Liu and Nandan Thakur and Crystina Zhang and Luyu Gao and Wenhu Chen and Jimmy Lin},
      year={2025},
      url={https://arxiv.org/abs/2508.06600}
}

@misc{turkle2025,
  author       = {{HLT-COE@JHU}},
  title        = {Turkle: An open-source clone of Amazon Mechanical Turk},
  year         = 2025,
  howpublished = {\url{https://github.com/hltcoe/turkle}}
}

@article{team2025kimi,
  title={Kimi k2: Open agentic intelligence},
  author={Team, Kimi and Bai, Yifan and Bao, Yiping and Chen, Guanduo and Chen, Jiahao and Chen, Ningxin and Chen, Ruijue and Chen, Yanru and Chen, Yuankun and Chen, Yutian and others},
  journal={arXiv preprint arXiv:2507.20534},
  year={2025}
}

@article{scikit-learn,
    title={Scikit-learn: Machine Learning in {P}ython},
    author={Pedregosa, F. and Varoquaux, G. and Gramfort, A. and Michel, V.
            and Thirion, B. and Grisel, O. and Blondel, M. and Prettenhofer, P.
            and Weiss, R. and Dubourg, V. and Vanderplas, J. and Passos, A. and
            Cournapeau, D. and Brucher, M. and Perrot, M. and Duchesnay, E.},
    journal={Journal of Machine Learning Research},
    volume={12},
    pages={2825--2830},
    year={2011}
}

@article{SALTON1988513,
title = {Term-weighting approaches in automatic text retrieval},
journal = {Information Processing \& Management},
year = {1988},
url = {https://www.sciencedirect.com/science/article/pii/0306457388900210},
author = {Gerard Salton and Christopher Buckley},
}

@misc{ wiki:xxx,
   author = "MediaWiki",
   title = "API:Search --- MediaWiki{,} ",
   year = "2024",
   url = "https://www.mediawiki.org/w/index.php?title=API:Search&oldid=6905053",
   note = "[Online; accessed 25-September-2025]"
 }
\bibliographystyle{iclr2026_conference}

\clearpage
\appendix

\addtocontents{toc}{\protect\setcounter{tocdepth}{3}}

\hypersetup{linkcolor=black}

\renewcommand{\contentsname}{\large Appendix Contents}
\let\oldtableofcontents\tableofcontents

\renewcommand{\tableofcontents}{%
    \begingroup
    \setlength{\cftbeforesubsecskip}{2pt} 
    \hypertarget{toc}{}
    \oldtableofcontents
    \endgroup
}

\tableofcontents

\clearpage

\pagestyle{fancy}
\renewcommand{\headrulewidth}{0pt}
\fancyhead{}
\fancyfoot[R]{\hyperlink{toc}{Back to Appendix Table of Contents}}

\clearpage

\section{The Use of Large Language Models}

In this work, large language models (LLMs) are employed solely as tools for data generation, as described in the main paper. Importantly, no parts of the manuscript are generated by LLMs. Hence, there are no concerns of plagiarism or scientific misconduct related to text generation.

\section{Dataset Construction: Knowledge}
\label{appendix knowledge}


\subsection{Annotator Biography}

The detailed biographies of the annotators involved in \ours construction are presented in \textbf{\autoref{tab:annotators}}. All annotators are from universities ranked in the Top 500 of the 2025 QS Global Rankings\textsuperscript{3} and are fluent in English. Annotators assess document–query relevance by judging whether a document facilitates answering the query. To ensure quality, independent validators conduct an additional round of verification.

\begin{table}[ht]
\caption{Biographies of 24 annotators involved in \ours construction (Author biographies are hidden to protect identity confidentiality).}
\label{tab:annotators}
\centering
\begin{adjustbox}{width=\textwidth,center}
\begin{tabular}{cl>{\raggedright\arraybackslash}p{3.5cm}>{\raggedright\arraybackslash}p{4.5cm}cc}
\toprule
\textbf{ID} & \textbf{Year} & \textbf{Major} & \textbf{Assigned Subject(s)} & \textbf{Author?} & \textbf{Validator?} \\
\midrule
1 & 3rd year Undergraduate & Biological Engineering & Biology & \xmark & \xmark \\
2 & 1st year Master & Biological Engineering & Biology & \xmark & \cmark \\
3 & 1st year Master & Biomedical Engineering & Biology, Pharmacy & \xmark & \xmark \\
4 & 2nd year Master & Biomedical Engineering & Biology, Pharmacy & \xmark & \xmark \\
5 & 1st year Master & Biomedical Engineering & Biology, Pharmacy & \xmark & \xmark \\
6 & 1st year PhD & Chemistry & Chemistry & \xmark & \xmark \\
7 & 2nd year Master & Chemistry & Chemistry & \xmark & \cmark \\
8 & 3rd year PhD & Medicine & Basic Medicine & \xmark & \xmark \\
9 & 3rd year Undergraduate & Clinical Medicine & Clinical Medicine, Diagnostics & \xmark & \xmark \\
10 & 3rd year Undergraduate & Medicine & Basic Medicine  & \xmark & \cmark \\
11 & 2nd year Master & Clinical Medicine & Clinical Medicine, Diagnostics & \xmark & \xmark \\
12 & 2nd year Master & Clinical Medicine & Clinical Medicine, Diagnostics & \xmark & \cmark \\
13 & 3rd year Undergraduate & Pharmacology & Pharmacy & \xmark & \cmark \\
14 & 4th year Undergraduate & Pharmacology & Pharmacy & \xmark & \xmark \\
15 & 1st year Master & Music & Music & \xmark & \xmark \\
16 & 1st year Master & Clinical Medicine & Clinical Medicine & \xmark & \xmark \\
17 & 1st year PhD & Sociology & Sociology, Psychology & \xmark & \xmark \\
18 & 1st year Master & Bioinformatics & Biology & \xmark & \xmark \\
19 & 2nd year PhD & Agricultural and Biosystems Engineering & Agriculture & \xmark & \xmark \\
20 & 4th year Undergraduate & Literature & History, Literature & \xmark & \xmark \\
21 & 3rd year Undergraduate & Geography and Environmental Studies & Geography & \xmark & \xmark \\
22 & 4th year PhD & Computer Science & - & \cmark & \cmark \\
23 & 4th year Undergraduate & Computer Science & - & \cmark & \cmark \\
24 & 3rd year Undergraduate & Electronic Engineering & - & \cmark & \cmark \\
\bottomrule
\end{tabular}
\end{adjustbox}
\end{table}

\subsection{Annotation Guideline and Interface}

To facilitate data annotation, we develop the following interface based on Turkle ~\citep{turkle2025}, an open-source clone of Amazon’s Mechanical Turk. The annotation guideline and interface is detailed in \autoref{turkle1}, \autoref{turkle2}, and \autoref{turkle3}.

\begin{figure}[h]
    \centering
    \includegraphics[width=0.9\textwidth]{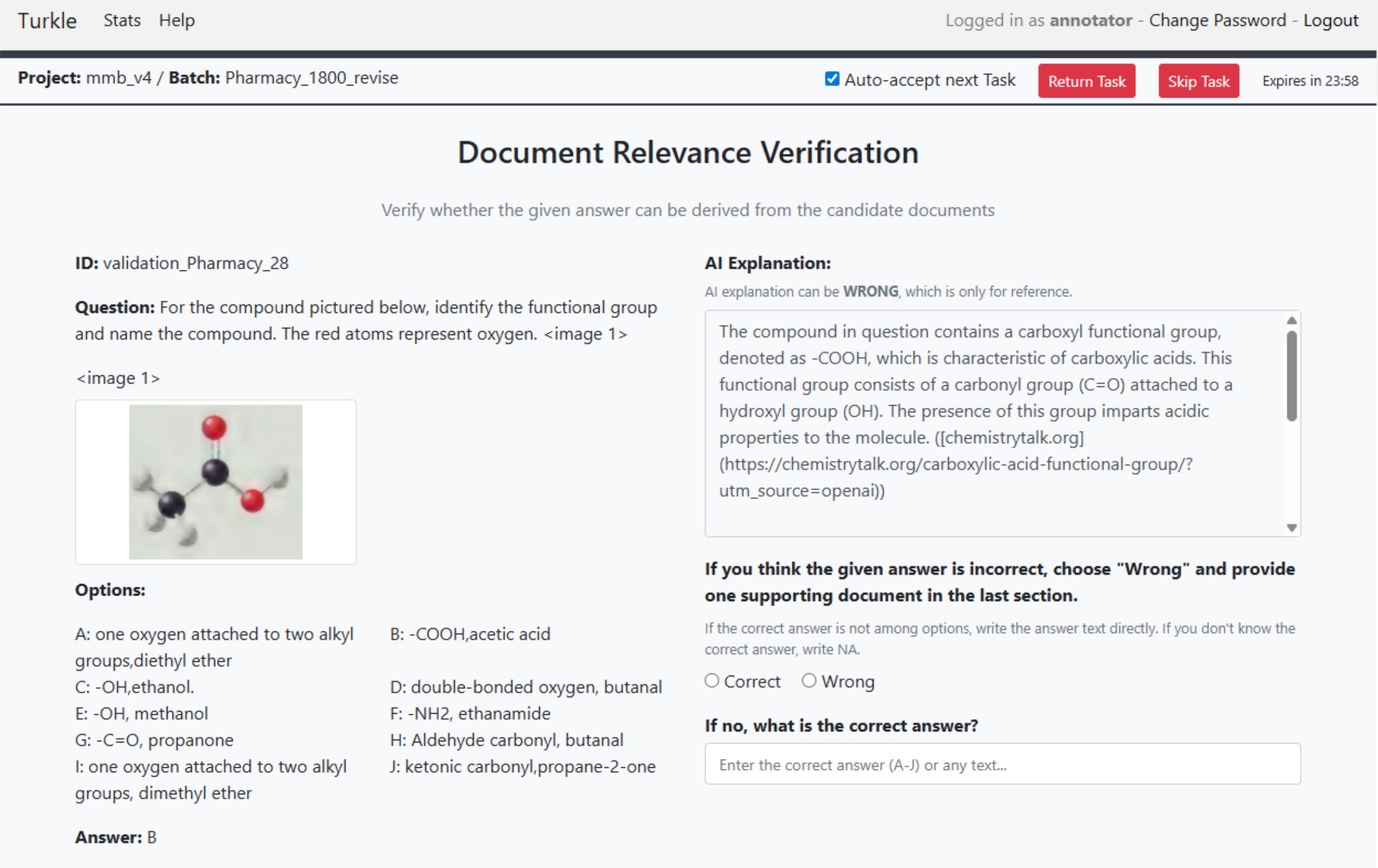}
    \caption{\textbf{Annotation Interface - Step 1: Question Understanding.} Annotators are first shown the question, associated images, candidate options, the correct answer, and an AI-generated explanation. The explanation is provided to aid understanding, though annotators are informed it may be incorrect. In this step, they judge whether the given answer is correct based on their own knowledge.}
    \label{turkle1}
\end{figure}

\begin{figure}[h]
    \centering
    \includegraphics[width=0.9\textwidth]{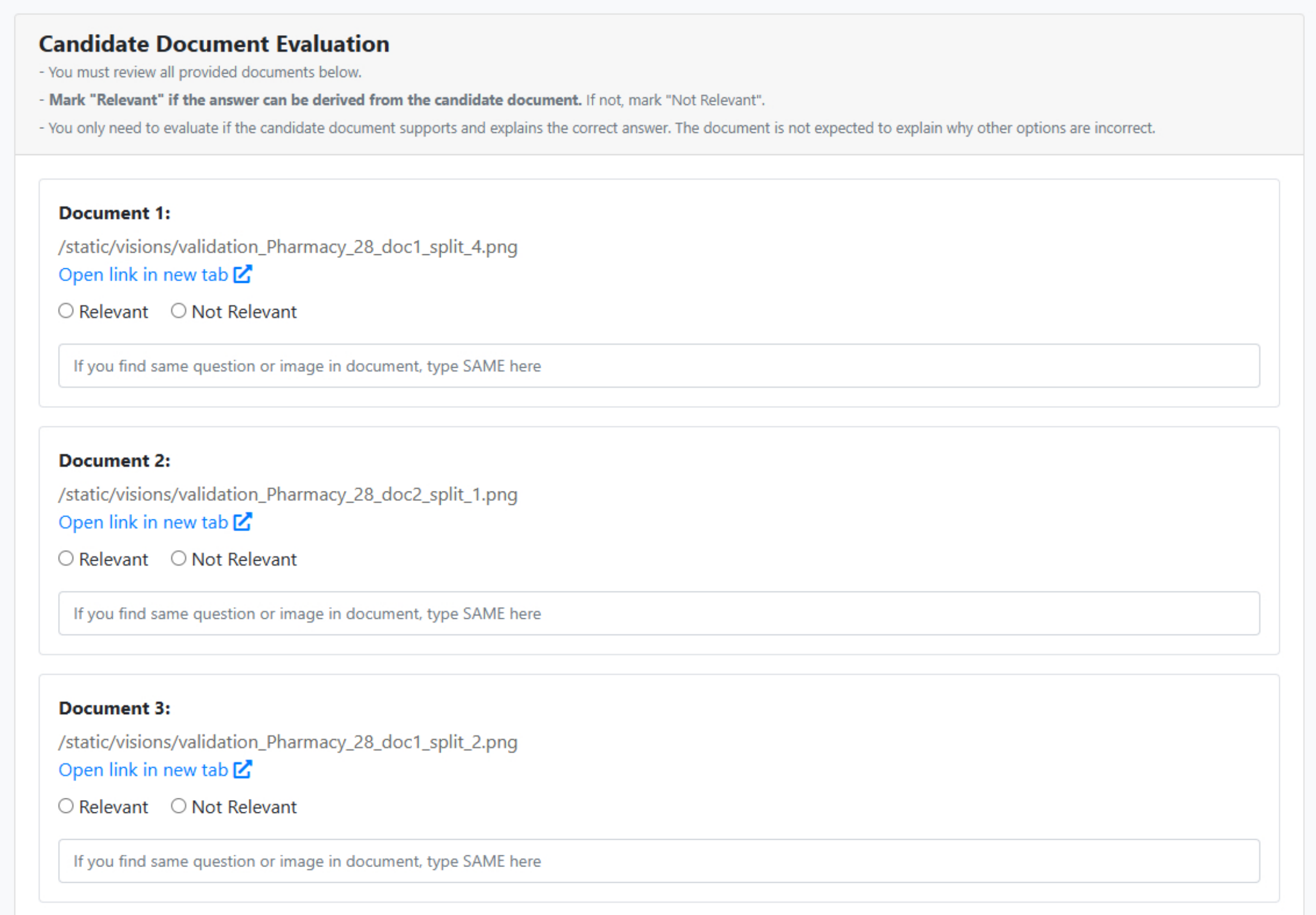}
    \caption{\textbf{Annotation Interface — Step 2: Candidate Document Evaluation.} After understanding the question, annotators are instructed to review candidate documents individually and judge whether each can facilitate correctly answering the question. Documents are shown in image format, with up to eight candidates presented. Document relevance definition has been explained to annotators before the annotation process.}
    \label{turkle2}
\end{figure}

\begin{figure}[h]
    \centering
    \includegraphics[width=0.9\textwidth]{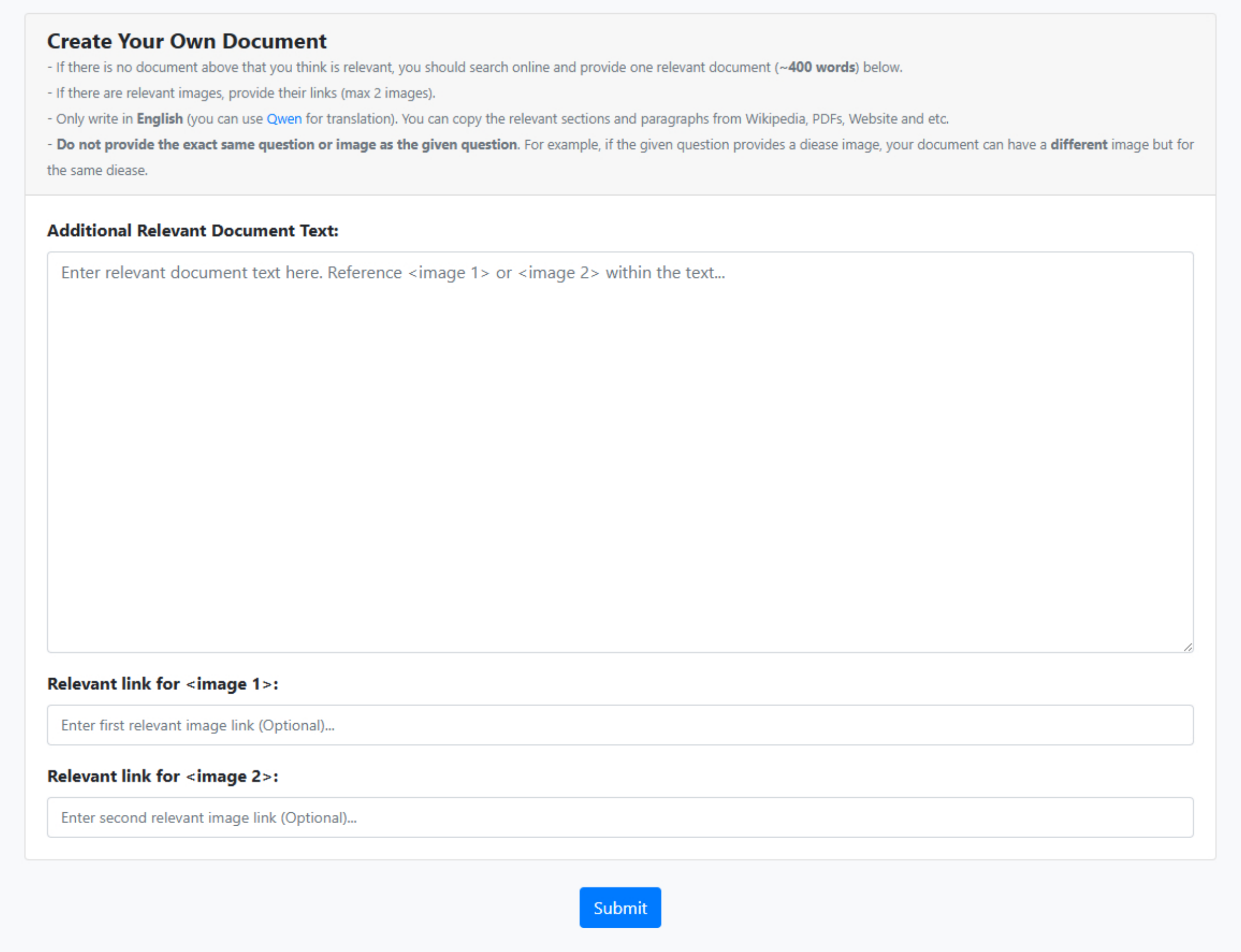}
    \caption{\textbf{Annotation Interface — Step 3: Create Relevant Document.} If none of the candidate documents are deemed relevant, annotators are required to search for a suitable web page and provide the gold evidence content. They are encouraged to include images from the source, and the final document is written in an interleaved image–text format.}
    \label{turkle3}
\end{figure}

\subsection{Data Annotation Payment}

The annotation and validation process for \ours spanned three months. Each annotator was assigned approximately \textbf{50 questions} aligned with their academic major. After annotation, validators independently assessed the quality of the labels. We provided a \textit{base rate} of \textbf{7 USD per hour}, with a quality adjustment of about 10\%. On average, annotating a single question required \textbf{10 minutes}, while validation took \textbf{4 minutes}. This compensation scheme ensured that annotators received wages competitive with the average teaching assistant salary at their universities. To maintain a manageable workload and reduce pressure, we recommended a maximum of \textbf{10 questions per day}.

\subsection{Dataset Construction Prompts}

The dataset construction prompts are presented in \Cref{tab:prompt_knowledge}, \Cref{tab:prompt_theorem}, \Cref{tab:prompt_search}, and \Cref{tab:prompt_relevance}.

\begin{figure}[htbp]
\centering
\begin{tcolorbox}[
    colback=black!7.5!white,
    colframe=black!30!white,
    fontupper=\footnotesize,
    fonttitle=\footnotesize,
    boxsep=5pt,
    left=5pt,
    right=5pt,
    top=5pt,
    bottom=5pt
]
Your task is to determine whether a question with images requires expert knowledge, such as about a historical event, scientific concept, economic theory, or medical disease. The last line of your response should be of the following format: ``Result: YES\_OR\_NO'' (without quotes). If the answer can be obtained easily by reading the question text and image content alone without the need of expert knowledge, say NO. Think step by step before answering. Here are some examples:

\begin{verbatim}
{example_1}
\end{verbatim}

\begin{verbatim}
{example_2}
\end{verbatim}

Now please determine whether this new question requires expert knowledge:
\begin{verbatim}
{question_and_answer}
\end{verbatim}

\end{tcolorbox}
\caption{The prompt for determining whether the question is knowledge-based.}
\label{tab:prompt_knowledge}
\end{figure}

\begin{figure}[htbp]
\centering
\begin{tcolorbox}[
    colback=black!7.5!white,
    colframe=black!30!white,
    fontupper=\footnotesize,
    fonttitle=\footnotesize,
    boxsep=5pt,
    left=5pt,
    right=5pt,
    top=5pt,
    bottom=5pt
]
Is the multimodal question testing a theorem, formula, equation, or algorithm in domains such as physics, economics, finance, computer science, and math? Answer YES or NO directly.

\begin{verbatim}
{question_and_answer}
\end{verbatim}

\end{tcolorbox}
\caption{The prompt for determining whether the question is theorem-based.}
\label{tab:prompt_theorem}
\end{figure}

\begin{figure}[htbp]
\centering
\begin{tcolorbox}[
    colback=black!7.5!white,
    colframe=black!30!white,
    fontupper=\footnotesize,
    fonttitle=\footnotesize,
    boxsep=5pt,
    left=5pt,
    right=5pt,
    top=5pt,
    bottom=5pt
]
Explain the answer to the question in a clear and detailed manner. Include citation web links to support the explanation — use relevant Wikipedia pages whenever possible. If a Wikipedia page is not available, use other reliable sources.

\begin{verbatim}
{question_and_answer}
\end{verbatim}

\end{tcolorbox}
\caption{The prompt for searching relevant web pages using GPT-Search.}
\label{tab:prompt_search}
\end{figure}

\begin{figure}[htbp]
\centering
\begin{tcolorbox}[
    colback=black!7.5!white,
    colframe=black!30!white,
    fontupper=\footnotesize,
    fonttitle=\footnotesize,
    boxsep=5pt,
    left=5pt,
    right=5pt,
    top=5pt,
    bottom=5pt
]

Question:

\begin{verbatim}
{question_and_answer}
\end{verbatim}

Document:

\begin{verbatim}
{document}
\end{verbatim}

You are a document analysis assistant. Your task is to determine whether the given document above answers the given question and supports the given answer.

\textbf{Instructions:}
\begin{enumerate}
    \item If the answer can be derived or inferred from the text and images in the document, respond YES; otherwise, respond NO.
    \item If the document discusses related topics but does not directly answer the given question, respond NO.
    \item If the document only provides reference paper titles without substantive content that supports the answer, respond NO.
\end{enumerate}

First, think step by step and explain your reasoning. In the last separate line, directly respond YES or NO without quotes.

\end{tcolorbox}
\caption{The prompt for judging whether the document is relevant to the question.}
\label{tab:prompt_relevance}
\end{figure}


\section{Dataset Construction: Theorem}
\label{appendix theorem}

\subsection{Theorem Database Construction}

The BRIGHT theorem corpus was embedded using Qwen3-Embedding \cite{qwen3embedding} and indexed in ChromaDB, which supports efficient semantic search via HNSW \cite{chromadb}. Each entry retains a unique \texttt{theorem\_id} and the original \texttt{text}, enabling fast, semantics-aware retrieval with full traceability to the source.

\subsection{Wikipedia Content Processing Pipeline}

We retrieved Wikipedia content by querying the MediaWiki Search API \cite{wiki:xxx} using theorem names as search keys. For supplementary sources in PDF format, we employed MonkeyOCR \cite{monkeyocr} to convert scanned documents into Markdown. The resulting text was then processed through a structured extraction prompt (Figure~\ref{tab:prompt_theorem_cleaning}) using GPT-5 to perform final cleaning, normalization, and precise theorem statement extraction.




\begin{figure}[htbp]
\centering
\begin{tcolorbox}[
    colback=black!7.5!white,
    colframe=black!30!white,
    fontupper=\footnotesize,
    fonttitle=\footnotesize,
    boxsep=5pt,
    left=5pt,
    right=5pt,
    top=5pt,
    bottom=5pt
]
You are given a markdown document. Your task is to extract the specific theorem, formula, equation, algorithm, or concept named ``\texttt{\{theorem\_name\}}'' from this document.

\textbf{Instructions:}
\begin{enumerate}
    \item Carefully locate the section that describes the theorem ``\texttt{\{theorem\_name\}}''.
    \item Extract the complete definition, explanation, and any associated formulas or equations.
    \item Remove all reference citations.
    \item If there are referenced images in the content, preserve the image references exactly as they appear.
    \item Your response MUST follow the following LaTeX-style format:
    
    \begin{verbatim}
\begin{definition}[{theorem_name}]
Complete definition and explanation, 
preserving mathematical notation.
Include examples if present.
\end{definition}
    \end{verbatim}
\end{enumerate}

Here is the document content:\\
\texttt{\{markdown\_content\}}
\end{tcolorbox}
\caption{The prompt for cleaning the theorem content.}
\label{tab:prompt_theorem_cleaning}
\end{figure}

\subsection{Document Deduplication}
\label{dedup}
All theorems extracted from Wikipedia were deduplicated prior to inclusion in the corpus. Deduplication was performed in two stages: first by theorem name, and then by semantic content using TF-IDF–based cosine similarity \cite{SALTON1988513}. Specifically, we employed \texttt{TfidfVectorizer} to compute \texttt{TF-IDF} vectors for all theorem statements \cite{scikit-learn}, followed by pairwise cosine similarity. Entries with near-identical content (cosine similarity $\geq 0.85$) were collapsed into a single representative instance.

\section{Dataset Construction: Contradiction}
\label{appendix contra}

\subsection{Negation}
\label{appendix negation}
First, we randomly select 200 samples from the COCO \cite{coco} dataset, each containing at least three positive objectives. For each entry, we construct a description using the template, ``The image includes $a$, $b$, $c$, but no $d$.'' In the positive description, we randomly select three positive objectives to replace $a$, $b$, and $c$, and select one negative objective to replace d. For the negative description, we generate two variations: one where all four objectives ($a$, $b$, $c$, $d$) are selected from the positive objectives, and another where one of $a$, $b$, or $c$ is replaced by $a$ randomly selected negative objective. The image from each sample is used as the query, and the three positive descriptions and one negative description are used as the corpus. Finally, we manually review the 200 queries and corresponding gold documents to ensure that the contradictory descriptions are identifiable by humans, and revise any ambiguous queries for clarity. No LLM prompting is involved in constructing the Negation task.

\subsection{Vehicle Design}
\label{appendix design}
On one hand, to construct the queries, we use design cases from the DesignQA dataset \cite{designqa} and augment them through appropriate modifications, such as altering numerical values and introducing variations in image elements. On the other hand, to construct the corpus, we apply MonkeyOCR \cite{monkeyocr} to extract and segment the Formula SAE Rulebook into 700 files, organized by rule ID. Finally, we review all the queries to ensure they represent incorrect designs.

\subsection{Traffic Case}
\label{appendix traffic}

First, we select a set of traffic rules and, based on these rules, create traffic violation cases by crafting relevant stories. These stories are then used as prompts to generate 12 images for each story using GPT-5. Afterwards, we manually review all the generated images and use Doubao \cite{doubao} to refine and enhance them for better clarity and relevance. Additionally, we leverage Doubao to generate specific objectives from the queries in order to construct image–text interleaved queries. For the corpus, we use MonkeyOCR to split Basic Theory of Driving and Final Theory of Driving \cite{sg_driving}, two official driving handbooks in Singapore, into separate files, which are then organized and used as the corpus. Finally, we conduct a manual review of all the queries, ensuring that any additional corpus IDs caused by excessive image details are properly incorporated into the queries.

\section{Experiment Details}
\subsection{Models and Instructions}
\label{experiment}

\begin{table}[ht]
\caption{Details of the multimodal retriever models evaluated in \ours.}
\label{tab:retrievers}
\centering
\begin{adjustbox}{width=0.9\textwidth,center}
\begin{tabular}{lll}
\toprule
\textbf{Model} & \textbf{Size} & \textbf{Version} \\
\midrule
BGE-M3 \cite{bge_m3} & 600M & BAAI/bge-m3 \\
NE-Embed-V2 \cite{nv_embed} & 8B & nvidia/NV-Embed-v2 \\
Qwen3-Embedding \cite{qwen3embedding} & 8B & Qwen/Qwen3-Embedding-8B \\
\midrule
EVA-CLIP \cite{evaclip} & 400M & QuanSun/EVA02-CLIP-L-14 \\
SigLIP \cite{siglip} & 650M & google/siglip-large-patch16-256 \\
JinaCLIP \cite{jinaclip} & 860M & jinaai/jina-clip-v2  \\
OpenCLIP \cite{openclip} & 1.4B & laion/CLIP-ViT-g-14-laion2B-s34B-b88K \\

\midrule
VISTA \cite{vista} & 200M & BAAI/bge-visualized-m3 \\
VLM2Vec \cite{vlm2vec} & 4B & TIGER-Lab/VLM2Vec-Full \\
GME-Qwen2-VL \cite{gme} & 7B & Alibaba-NLP/gme-Qwen2-VL-7B-Instruct \\
Ops-MM-Embedding \cite{ops} & 7B & OpenSearch-AI/Ops-MM-embedding-v1-7B  \\
E5-V \cite{e5v} & 8B & royokong/e5-v \\
MM-Embed \cite{mmembed} & 8B & nvidia/MM-Embed \\
\midrule
ColPali \cite{colpali} & 3B & vidore/colpali-v1.3 \\
\bottomrule
\end{tabular}
\end{adjustbox}
\end{table}

Following \tiir, we evaluate text retrievers on multimodal retrieval tasks by replacing images with captions generated by an LLM. To simulate real-time inference, we apply the standardized prompt ``Describe the image" and use Qwen2-VL-2B-Instruct to produce the captions.

\renewcommand{\arraystretch}{1.2} 
\begin{table}[ht]
\caption{Instruction prompts used during model evaluation in \ours.}
\label{tab:instruction}
\centering
\begin{adjustbox}{width=1\textwidth,center}
\begin{tabular}{lll}
\toprule
\textbf{Task} & \textbf{Modality} & \textbf{Prompt} \\
\midrule
\multirow{2}{*}{Knowledge}  & Multimodal & \multirow{2}{*}{Retrieve relevant documents that help answer the question.} \\
& Text &  \\
\midrule
\multirow{2}{*}{Theorem}  & Multimodal & \multirow{2}{*}{Retrieve relevant theorems that are involved in solving the problem.} \\
& Text &  \\
\midrule
\multirow{2}{*}{Negation} & Multimodal & Given an image, retrieve descriptions that have contradictory information with the image. \\
 & Text & Given an image caption, retrieve descriptions that have contradictory information with the image caption. \\
\midrule
\multirow{2}{*}{Vehicle Design} & Multimodal & Given a vehicle design, retrieve the design requirements that it violates. \\
& Text & Given a vehicle design description, retrieve the design requirements that it violates. \\
\midrule
\multirow{2}{*}{Traffic Case} & Multimodal & Given a traffic case, retrieve the driving rule documents that it violates. \\
& Text & Given a traffic case description, retrieve the driving rule documents that it violates.\\
\bottomrule
\end{tabular}
\end{adjustbox}
\end{table}

\subsection{Implementations and Machines}
The \ours dataset is constructed following the conventions of MTEB \cite{mteb}, including data format and evaluation pipeline, with modifications to support mixed-modality inputs during evaluation. All experiments are conducted on NVIDIA A100, A6000, or H100 GPUs. The runtime of a full evaluation depends on the model, but with the limited corpus size for efficiency, one complete run can be completed within 4 hours on a single A100 GPU for open-source dense models. To further accelerate dense model evaluation, we employ FlashAttention \cite{dao2022flashattention}.

\subsection{Detailed Results}

\begin{table}[htbp]
\centering
\caption{Detailed performance of retrieval models on \ours (\knowledge).}
\label{tab:breakdown_results}
\begin{adjustbox}{width=\textwidth,center}
\begin{tabular}{lcccccccccccccccccc}
\toprule
 & \multicolumn{16}{c}{Knowledge} & Avg. \\
\cmidrule(lr){2-17} 
Model & Music & Design & Theo. & Art & Hist. & Soci. & Psy. & Lit.& Pharm. & Diag. & Clinic. & Basic. & Agri. & Geo. & Chem. & Bio. & \\
\midrule
\multicolumn{16}{c}{\textit{Text Models with Image Caption}} \\
\midrule
BGE-M3 & 43.4 & 44.0 & 49.4 & 57.2 & 47.7 & 39.5 & 52.2 &15.8 & 58.5 &11.2&28.2& 36.2& 38.7 & 48.6 & 37.6 & 48.3 & 41.0 \\
NV-Embed-v2 & 63.8 & 61.8  & 70.1 & 86.8 & 70.6 & 64.3 & 59.7 & 95.8  & 78.0 & 19.8  & 46.0  & 59.0  &  65.3 & 63.3  & 70.0 & 63.6 & 64.9 \\
Qwen3-Embedding & 62.8 & 62.1 & 74.8 & 87.3 & 76.1 & 74.0 & 69.3 & 97.8 & 83.1 & 34.8 & 47.0 & 64.0 & 69.5 & 76.5 & 74.0 & 72.6 & 70.4 \\
\midrule
\multicolumn{16}{c}{\textit{Text and Image Two-Stream Models with Vector Fusion}} \\
\midrule
EVA-CLIP & 30.5 & 1.5 & 3.5 & 7.5& 16.7 & 5.5 & 16.3 & 0.0& 22.7& 10.3&10.0 & 16.4 &  41.6 & 15.4&  20.4 &  18.5 & 14.8 \\
SigLIP & 25.0 & 25.6 & 26.2 & 30.0 &  16.7 & 1.4 & 14.7 & 22.7 & 13.8 & 9.7& 15.6 & 19.6 & 30.2&18.3& 26.7 & 27.3 & 20.2\\
OpenCLIP & 20.9 & 50.7 &  62.9 & 86.4 & 35.8 & 10.2 & 15.1 & 22.7 & 11.1 & 10.6 & 20.8 & 25.8 &  34.1 & 45.8 & 23.9 & 34.3 & 31.9 \\
JinaCLIP & 18.5 & 11.0 & 23.0 & 33.1 & 14.2 & 0.0 & 17.1 & 0.0 & 17.8 & 6.1 & 21.7 & 21.1 & 35.1 & 24.7 & 30.4 & 15.4 & 18.1 \\
\midrule
\multicolumn{16}{c}{\textit{Multimodal Models with Merged Image}} \\
\midrule
VISTA & 39.3 & 3.5 & 17.2 & 27.5 & 12.3 & 13.9 & 28.0 & 0.0 & 48.9 & 18.2 & 23.9 & 31.2 & 33.6 & 22.0 & 36.9 & 33.1 & 24.3\\
E5-V & 13.0 & 23.4 & 17.6 & 46.1 & 15.6 & 4.3 & 10.8 & 7.7 & 12.5 & 7.1 & 13.5 & 13.7 & 18.3 & 13.1 & 23.3 & 10.0 & 15.6 \\
MM-Embed & 51.6 & 60.8 & 68.3 & 80.5 & 57.5 & 69.4 & 59.5 & 94.1 & 63.8 & 35.1 & 50.9 & 68.9 & 60.9 & 76.0 & 62.1 & 60.7 & 63.8 \\
VLM2Vec & 34.4 & 44.0 & 49.6 & 84.8 & 36.4 & 12.3 & 19.3 & 19.2 & 17.4 & 13.6 & 23.7 & 33.1 & 39.0 & 40.7 & 37.5 & 30.8 & 33.5 \\
GME-Qwen2-VL & 55.1 & 40.4 & 57.1 & 64.8 & 39.2 & 50.6 & 51.1 & 32.9 & 57.2 & 20.6 & 32.1 & 62.2 & 38.9 & 48.4 & 63.6 & 39.6 & 47.1 \\
Ops-MM-Embedding & 58.5 & 75.6 & 84.2 & 96.8 & 71.4 & 71.1 & 59.7  &  73.7 & 76.1&30.9& 50.7& 64.5 &58.7 & 78.5 &  80.4 & 69.0 & 68.7 \\
\midrule
\multicolumn{16}{c}{\textit{Multimodal Models with Document as Image}} \\
\midrule
GME-Qwen2-VL & 58.2 & 46.5 & 53.6 & 58.4 & 52.5 & 48.5  & 48.2  & 52.1 & 72.9 & 16.8&31.7& 40.2 & 49.7 & 69.0 &  53.8 & 45.4 & 49.8 \\
Ops-MM-Embedding & 60.6 & 59.0 & 68.4 & 82.4 &  68.3 & 63.0 & 58.6 & 68.3 & 74.3 & 31.2 &39.3 & 65.9 & 57.2 & 69.3 & 76.1 & 71.9  & 63.4 \\
ColPali & 25.1 & 27.7 & 46.4 & 43.7& 31.7 & 19.4 & 38.5  & 0.0 & 64.1 & 10.6& 23.0 & 60.1& 36.7 & 32.6 &  67.6 & 56.3 & 36.5 \\
\bottomrule
\end{tabular}
\end{adjustbox}
\end{table}

\clearpage

\section{Analysis Details}
\label{cases}

\subsection{False Positive and False Negative Analysis}

Although queries and their relevant documents were carefully validated by human annotators, false positives and false negatives may still arise when aggregating documents across queries or sampling from external corpora. We explicitly instructed annotators and validators to identify similar or related queries and to cross-annotate documents accordingly. As a result, some queries share the same relevant documents.

To quantitatively assess the prevalence of such labeling errors, we conducted a human audit of the top-retrieved documents retrieved by the best-performing model Ops-MM-Embedding. As shown in Table~\ref{tab:audit_results}, the audit revealed zero false positives and a false negative rate of only 2.5\% for \knowledge\ tasks for sampled 120 documents. Similarly, for \theorem\ tasks, the combined error rate was minimal at approximately 5.8\%, comprising 3.3\% false negatives and 2.5\% false positives. These results suggest that label noise is insignificant, thereby supporting the reliability of the benchmark.

For \contradiction\ tasks, the dataset is relatively small and predominantly constructed manually by annotators and validators. Given the quality control in the construction process, no additional human evaluation was deemed necessary.

\begin{table}[htbp]
\centering
\caption{Human audit of document relevance annotations for the top-retrieved documents produced by the best-performing multimodal model, Ops-MM-Embedding. A \textit{false negative} is a relevant document incorrectly labeled as irrelevant by our method, and a \textit{false positive} is an irrelevant document incorrectly labeled as relevant.}
\label{tab:audit_results}
\begin{adjustbox}{width=0.75\textwidth,center}
\begin{tabular}{lccccc}
\toprule
 &  & \multicolumn{2}{c}{False Negatives} & \multicolumn{2}{c}{False Positives} \\
\cmidrule(lr){3-4} \cmidrule(lr){5-6} 
Dataset & Documents Checked & Count & Ratio & Count & Ratio \\
\midrule
\theorem & 120 & 4 & 3.3\% & 3 & 2.5\% \\
\knowledge & 120 & 3 & 2.5\% & 0 & 0.0\% \\
\bottomrule
\end{tabular}
\end{adjustbox}
\end{table}

\subsection{Error Case Studies}

In this section, we present case studies for the Ops-MM-Embedding model in different domains such as Biology (\Cref{case: fauna}) and Traffic Case (\Cref{case: tunnel}). The error case analysis for \theorem tasks are exemplified as follows.

A recurring issue in Engineering and Geometry tasks is the model's tendency to perform coarse-grained matching based on shape and keywords, while ignoring the specific geometric conditions or physical constraints defined in the query.

\textbf{Case Study: Engineering (Geometry)} \\
\textit{Query ID: validation\_Architecture\_and\_Engineering\_5}
\begin{itemize}
    \item \textbf{Query Content:} An image showing a pentagon with internal angles and a specific coordinate bearing angle ($\alpha_{12}=30^\circ$), asking to calculate other bearing angles.
    \item \textbf{Retrieved Negative (Top-1):} ``Inscribing Circle in Regular Pentagon''.
    \item \textbf{Analysis:} This represents a \textbf{keyword and shape hallucination}. The model correctly identifies the visual object (a pentagon) and the domain (geometry/angles). However, it retrieves a document about inscribing circles—likely because the dense geometric keywords and the visual of a polygon strongly correlate in the embedding space. The model fails to attend to the specific logical task (calculating bearing angles) and instead prioritizes the dominant visual features (the pentagon shape).
\end{itemize}


\textbf{Case Study: Engineering (Statics)} \\
\textit{Query ID: test\_Architecture\_and\_Engineering\_214}
\begin{itemize}
    \item \textbf{Query Content:} A floor plan asking to compute the ``tributary areas'' for a specific floor beam B1.
    \item \textbf{Retrieved Negative (Top-1):} ``Static equilibrium''.
    \item \textbf{Analysis:} The retrieved document discusses the static equilibrium of beams. While semantically related to the domain (structural engineering and beams), it is a conceptual mismatch. The model retrieves a theoretical concept (statical indeterminacy) rather than the procedural knowledge required for area calculation. This suggests that the retriever struggles to distinguish between \textit{theoretical concepts} and \textit{practical calculation tasks} when visual cues (schematic diagrams of beams) are similar.
\end{itemize}


In scientific domains, the model often exhibits ``partial understanding'', where it correctly identifies the strict sub-domain or topic but fails to retrieve the document addressing the specific variable or relationship queried.

\textbf{Case Study: Physics (Thermodynamics)} \\
\textit{Query ID: test\_Physics\_74}
\begin{itemize}
    \item \textbf{Query Content:} A $P-V$ (Pressure-Volume) graph showing a cyclic process, asking to identify the point of highest temperature.
    \item \textbf{Retrieved Negative (Top-1):} ``Isothermal process''.
    \item \textbf{Analysis:} The retrieved document explains isothermal processes (constant temperature), which frequently utilize $P-V$ diagrams similar to the query image. The retrieval is plausible but incorrect; the model latched onto the visual graph type ($P-V$ curve) but failed to deduce the specific relationship ($PV=nRT$) required to find the maximum temperature. This confirms the limitation regarding \textbf{higher-level deduction}: the model recognizes the graphical language but not the specific physical implication.
\end{itemize}


\textbf{Case Study: Math (Data Interpretation)} \\
\textit{Query ID: test\_Math\_469}
\begin{itemize}
\item \textbf{Query Content:} A histogram/bar chart showing student distances from school, asking for the percentage of students whose distance falls within the 5 km to 10 km range.
\item \textbf{Retrieved Negative (Top-1):} ``Generic statistical methods''.
\item \textbf{Analysis:} The retrieved document discusses general statistical techniques and data representation, which often involve histograms similar to the query image. The retrieval appears contextually related due to the presence of a histogram but is ultimately incorrect; the model associated the visual format with a broad statistical category but failed to extract or interpret the specific numerical data (bin counts and ranges) needed to compute the required percentage. This highlights a deficiency in \textbf{visual-numerical alignment}: the model recognizes the chart type but does not connect it to the mathmatical reasoning.
\end{itemize}

\begin{figure}[h]
    \centering
    \includegraphics[width=0.8\textwidth]{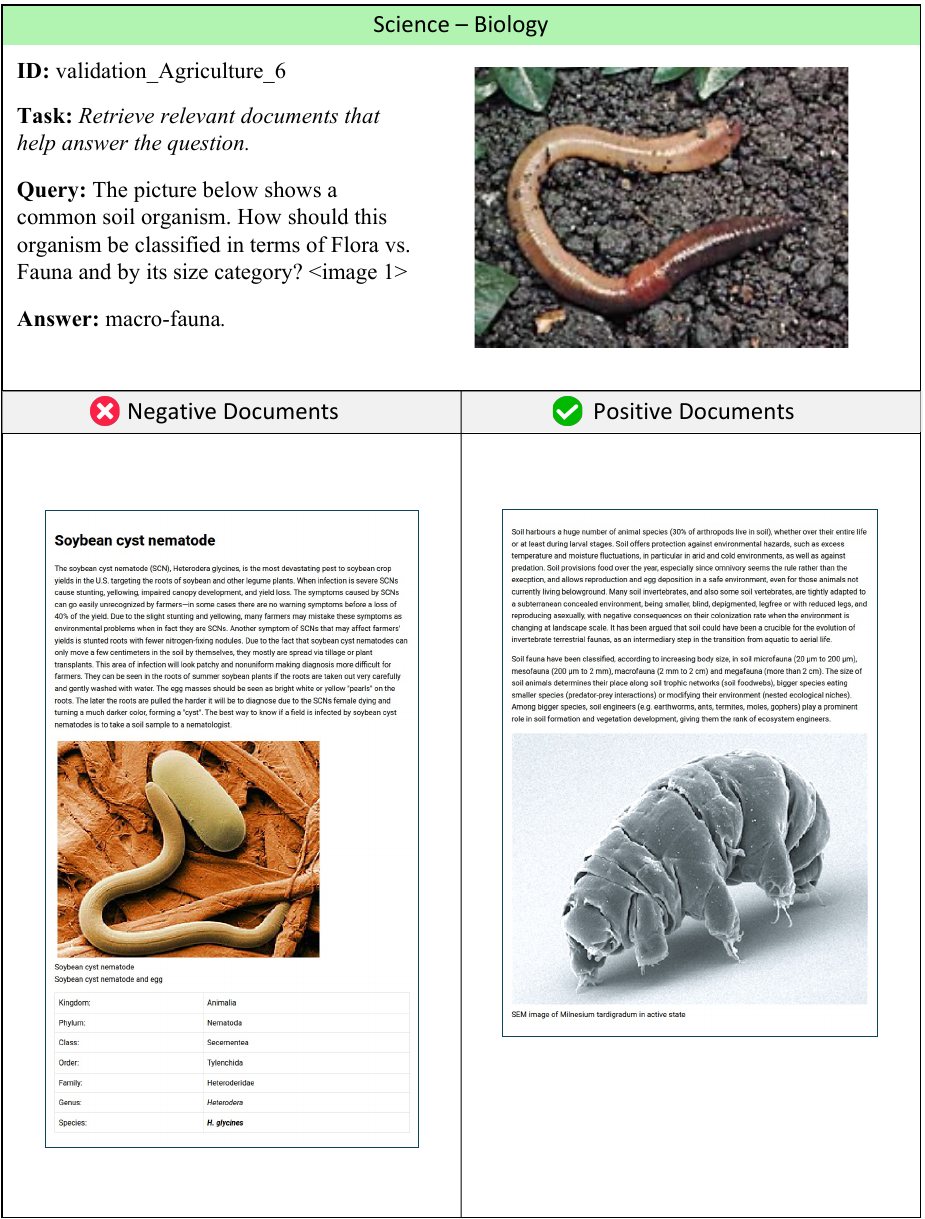}
    \caption{Error case example in Agriculture where the multimodal embedding model Ops-MM-Embedding prioritizes the negative document in the left over the positive document in the right.
    }
    \label{case: fauna}
\end{figure}

\begin{figure}[h]
    \centering
    \includegraphics[width=0.8\textwidth]{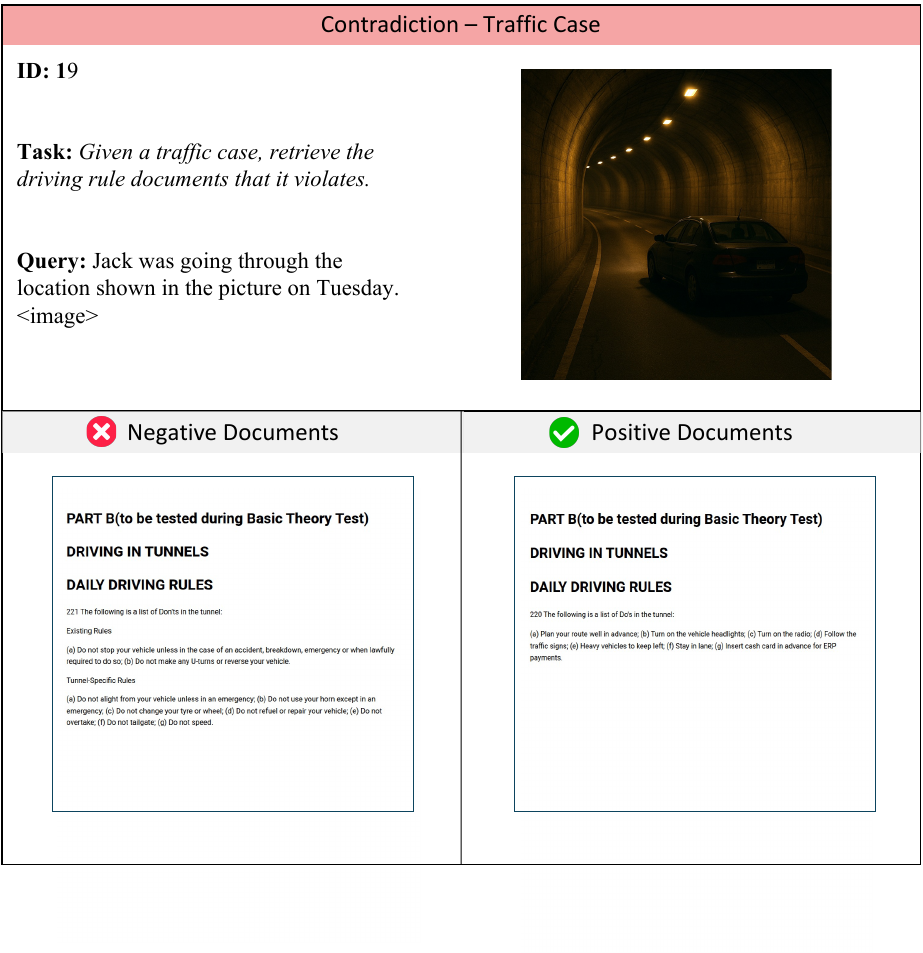}
    \caption{Error case example in Traffic where the multimodal embedding model Ops-MM-Embedding prioritizes the negative document in the left over the positive document in the right.
    }
    \label{case: tunnel}
\end{figure}

\clearpage
\subsection{Test-Time Scaling in Retrieval}
\label{test-time-sacling-in-retrieval}

We conducted query expansion experiments using both weak and strong vision-language models (VLMs)—namely, Qwen2-VL-2B and Qwen2.5-VL-72B—for weak and strong multimodal retrievers (i.e., GME-Qwen2-VL and Ops-MM-Embedding). As shown in \Cref{tab:test_scale,tab:test_scale_ops}, query expansion is generally effective for weak retriever models. However, for stronger retrievers, the quality of the query expansion becomes critical: expansions generated by the weaker VLM actually degrade the performance of the stronger retriever.

With query expansion by a strong LLM (Qwen2.5-VL-72B), the expansion technique is effective for improving both strong and weak retriever models. However, they are still far from perfect on this benchmark. For example, the best retriever with strong query expansion only achieves 55.9 for medical queries and 31.8 for math queries.

\begin{table}[htbp]
\centering
\caption{nDCG@10 scores of the multimodal retriever GME-Qwen2-VL on \ours \knowledge and \theorem tasks, comparing the original queries with query expansions generated by Qwen2-VL-2B-Instruct and Qwen2.5-VL-72B-Instruct. The average query length ($Q$ \#Text) before and after expansion is reported as the number of tokens measured by the GPT-2 tokenizer.}
\label{tab:test_scale}
\begin{adjustbox}{width=\textwidth,center}
\begin{tabular}{lccccccccccc}
\toprule
& \multicolumn{5}{c}{Knowledge} & \multicolumn{5}{c}{Theorem}  & Avg. \\
\cmidrule(lr){2-6} \cmidrule(lr){7-11} 
Model & $Q$ \#Text & Art & Med. & Sci. & Hum.  & $Q$ \#Text & Math & Phy. & Eng. & Bus.  & \\
\midrule
 
Original & 31.4 & 54.3& 40.1 & 46.8 & 45.6 & 58.6 & 28.8 & 36.0 & 30.2 & 45.1 & 40.9 \\
Qwen2-VL-2B & 699.6 & 64.9 & 49.6 & 64.6 & 48.9& 735.9 & 23.5 & 36.5 & 31.5 &  48.3 & 46.0 \\
Qwen2.5-VL-72B & 843.8 & 76.9 & 61.8 & 77.0 & 72.2 & 1218.4 & 33.3 & 36.9 & 32.7 & 55.0 & 55.7\\

\bottomrule
\end{tabular}
\end{adjustbox}
\end{table}

\begin{table}[htbp]
\centering
\caption{nDCG@10 scores of the multimodal retriever Ops-MM-Embedding on \ours \knowledge and \theorem tasks, comparing the original queries with query expansions generated by Qwen2-VL-2B-Instruct and Qwen2.5-VL-72B-Instruct. The average query length ($Q$ \#Text) before and after expansion is reported as the number of tokens measured by the GPT-2 tokenizer.}
\label{tab:test_scale_ops}
\begin{adjustbox}{width=\textwidth,center}
\begin{tabular}{lccccccccccc}
\toprule
& \multicolumn{5}{c}{Knowledge} & \multicolumn{5}{c}{Theorem}  & Avg. \\
\cmidrule(lr){2-6} \cmidrule(lr){7-11} 
Model & $Q$ \#Text & Art & Med. & Sci. & Hum.  & $Q$ \#Text & Math & Phy. & Eng. & Bus.  & \\
\midrule
 
Original &  31.4 & 79.3 & 52.5 & 70.0 & 67.8 & 58.6 & 27.7 &  39.5 & 30.1 & 52.3 & 52.4\\
Qwen2-VL-2B & 699.6 & 77.2 & 45.7 & 67.1 & 58.0 & 735.9 & 24.7 & 35.6 & 29.5 & 50.2 & 48.5 \\
Qwen2.5-VL-72B & 843.8 & 80.5 & 55.9 & 73.1 & 64.6 & 1218.4 & 31.8 & 39.4 & 35.3 & 53.9 & 54.3 \\

\bottomrule
\end{tabular}
\end{adjustbox}
\end{table}

\clearpage

\section{Data Examples}
\label{examples}

\begin{figure}[h]
    \centering
    \includegraphics[width=0.8\textwidth]{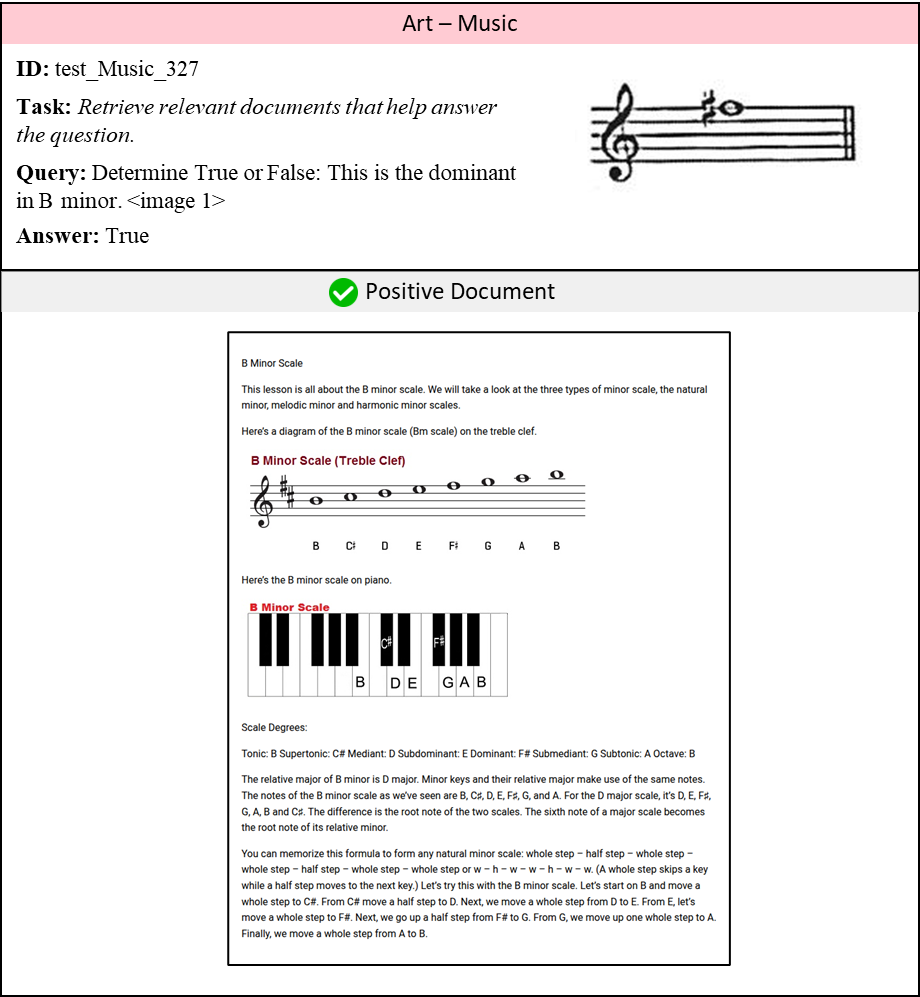}
    \caption{Music example.
    }
    \label{case: music}
\end{figure}

\begin{figure}[h]
    \centering
    \includegraphics[width=0.8\textwidth]{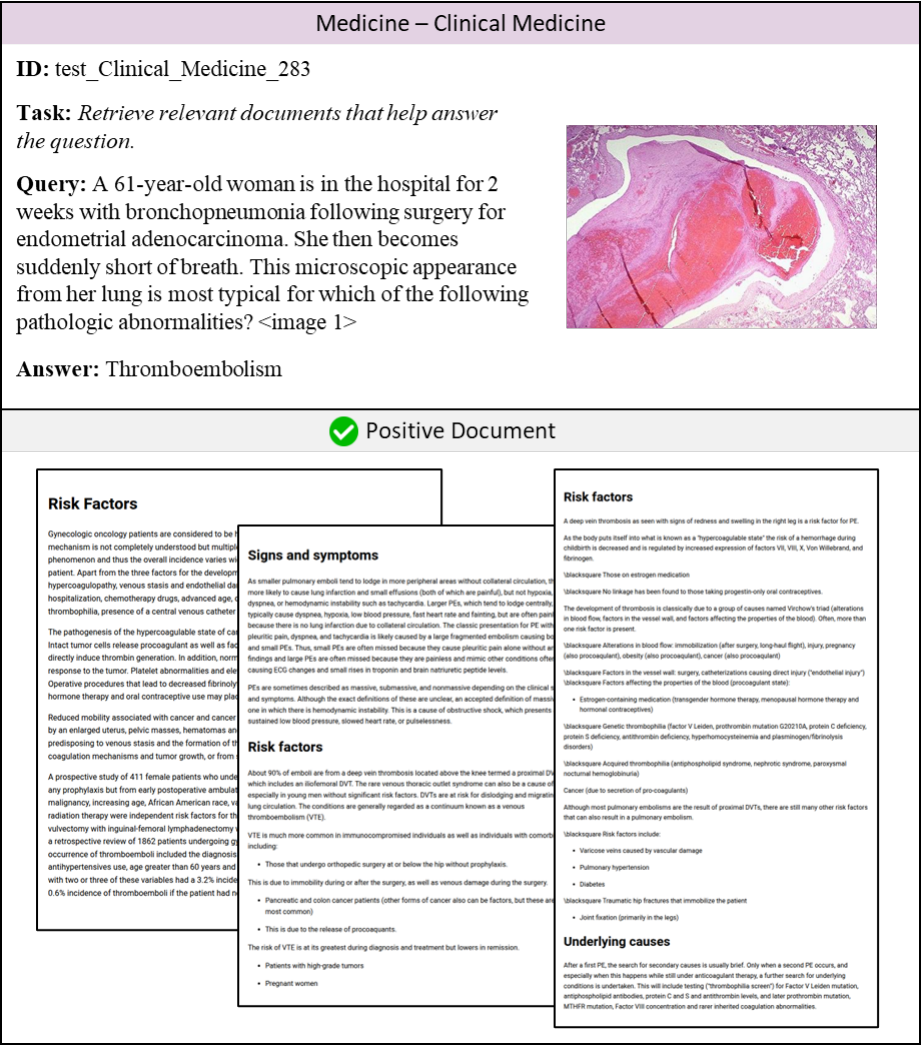}
    \caption{Clinic Medicine example.
    }
    \label{case: clinic}
\end{figure}

\begin{figure}[h]
    \centering
    \includegraphics[width=0.8\textwidth]{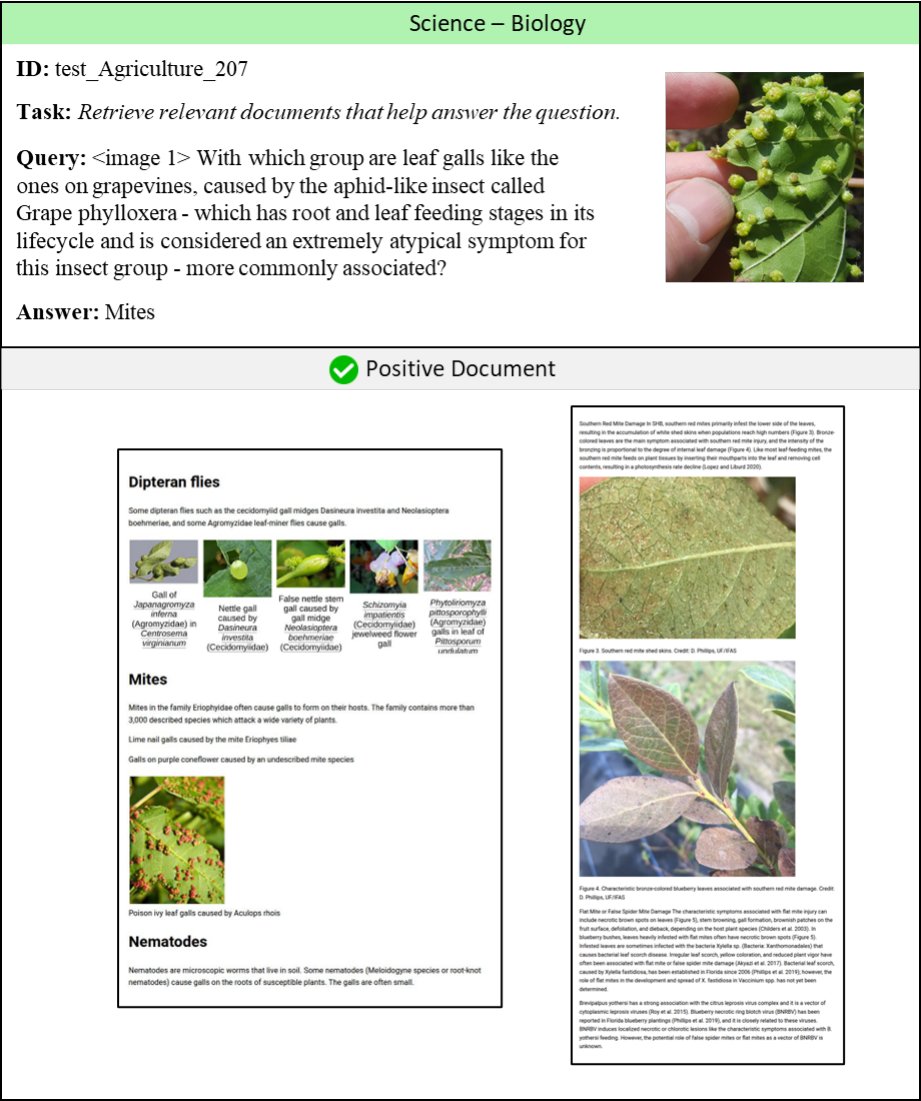}
    \caption{Biology example.
    }
    \label{case: biology}
\end{figure}

\begin{figure}[h]
    \centering
    \includegraphics[width=0.8\textwidth]{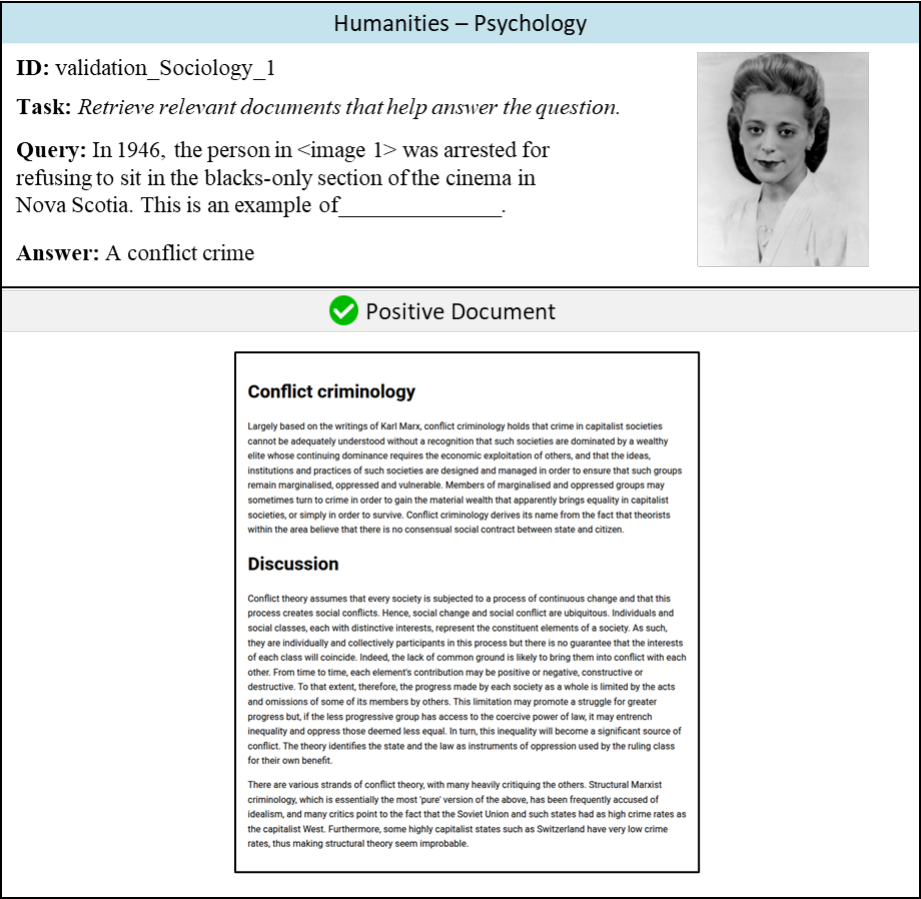}
    \caption{Psychology example.
    }
    \label{case: psychology}
\end{figure}


\begin{figure}[h]
    \centering
    \includegraphics[width=0.8\textwidth]{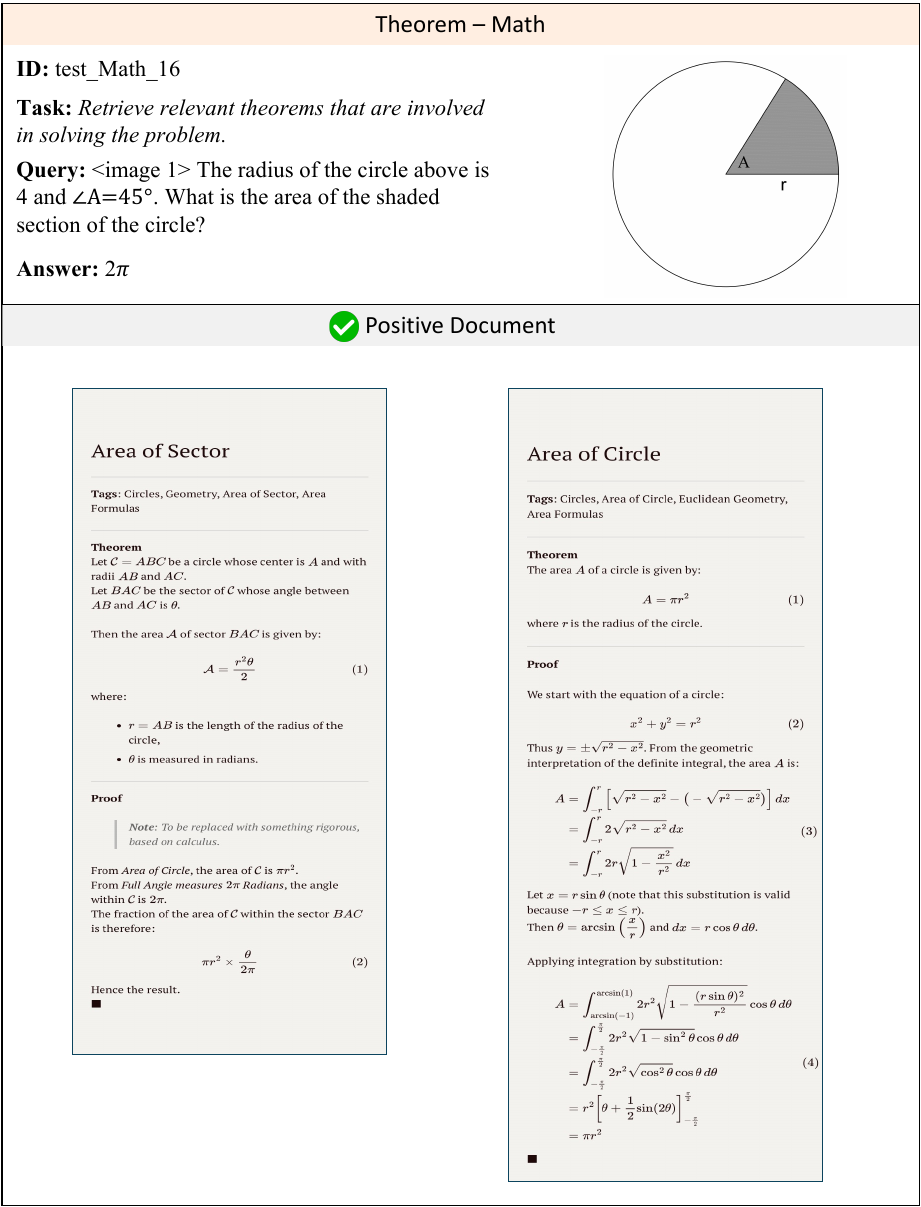}
    \caption{Math example.
    }
    \label{case: math}
\end{figure}

\begin{figure}[h]
    \centering
    \includegraphics[width=0.8\textwidth]{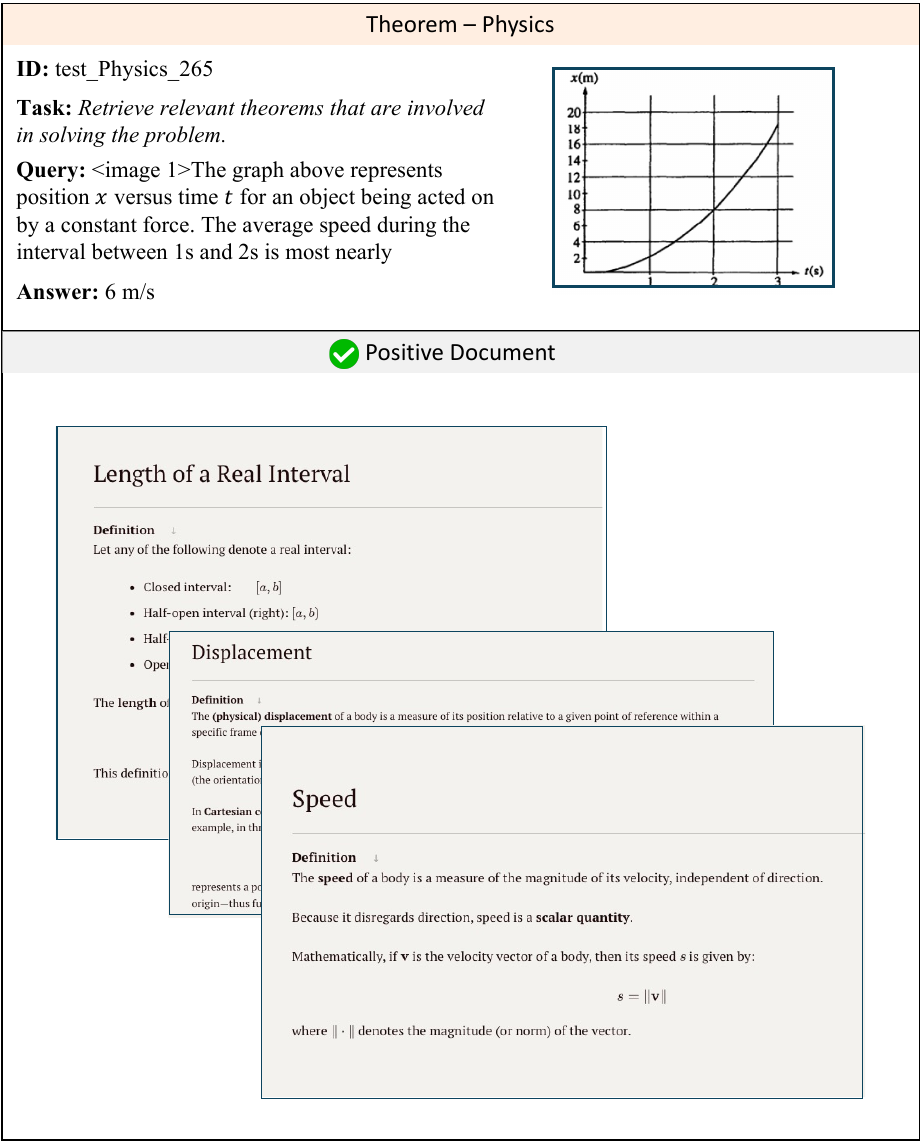}
    \caption{Physics example.
    }
    \label{case: physics}
\end{figure}

\begin{figure}[h]
    \centering
    \includegraphics[width=0.8\textwidth]{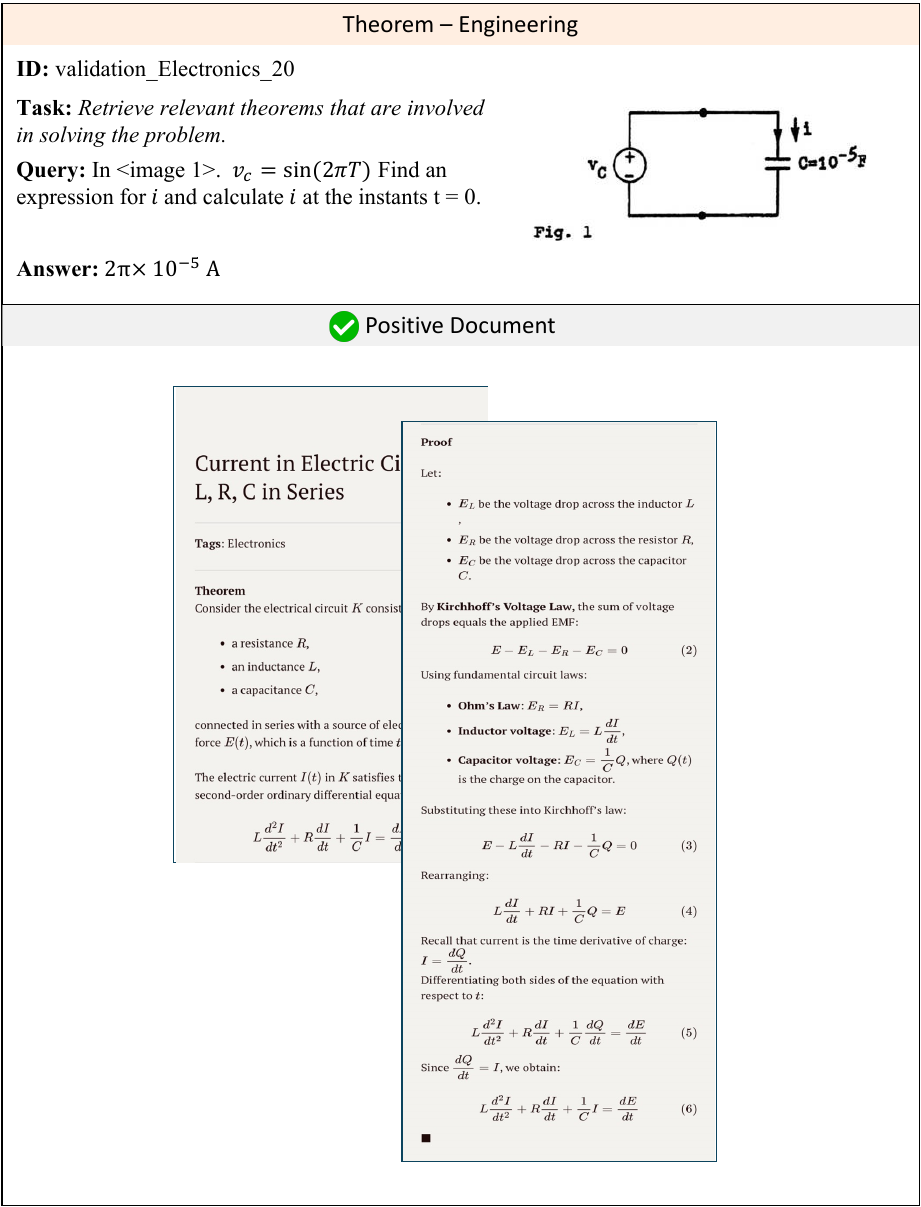}
    \caption{Engineering example.
    }
    \label{case: engineer}
\end{figure}

\begin{figure}[h]
    \centering
    \includegraphics[width=0.8\textwidth]{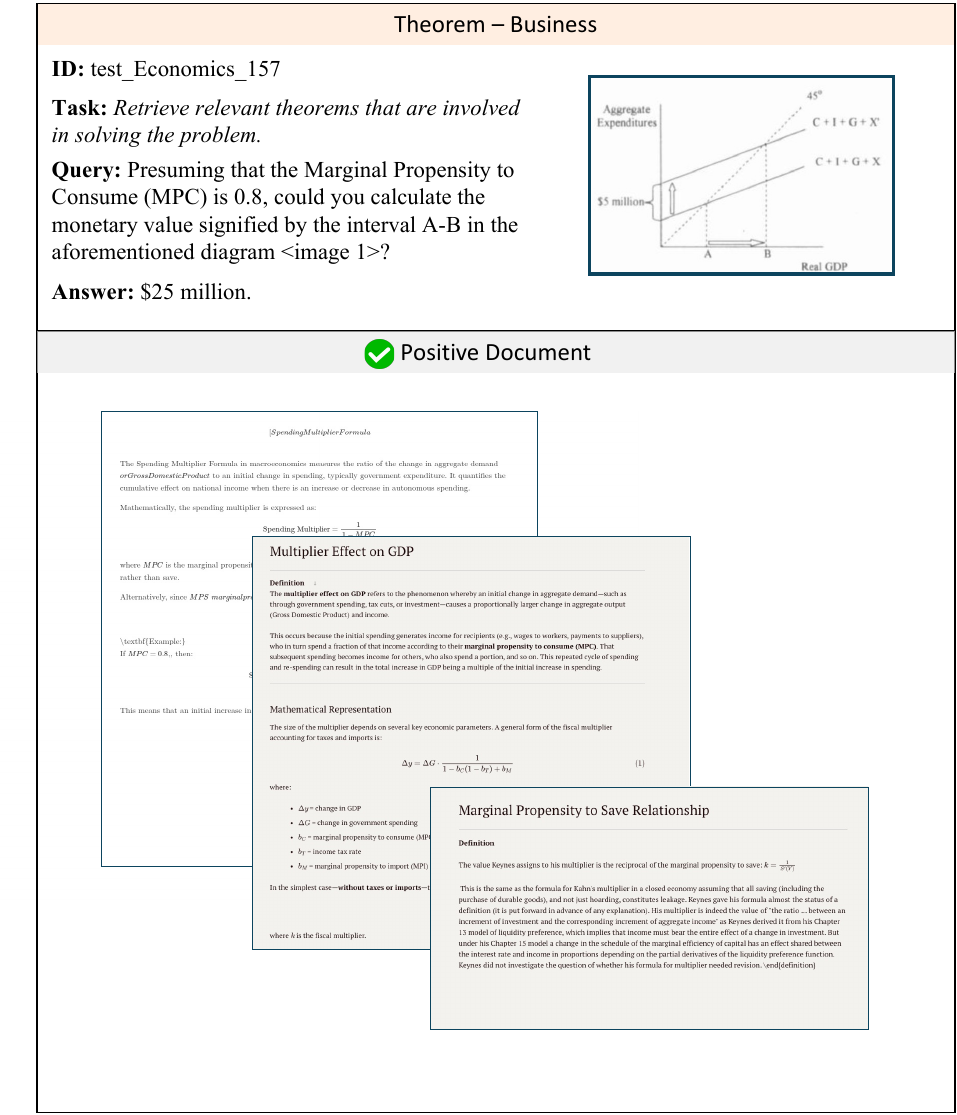}
    \caption{Business example.
    }
    \label{case: business}
\end{figure}

\begin{figure}[h]
    \centering
    \includegraphics[width=0.8\textwidth]{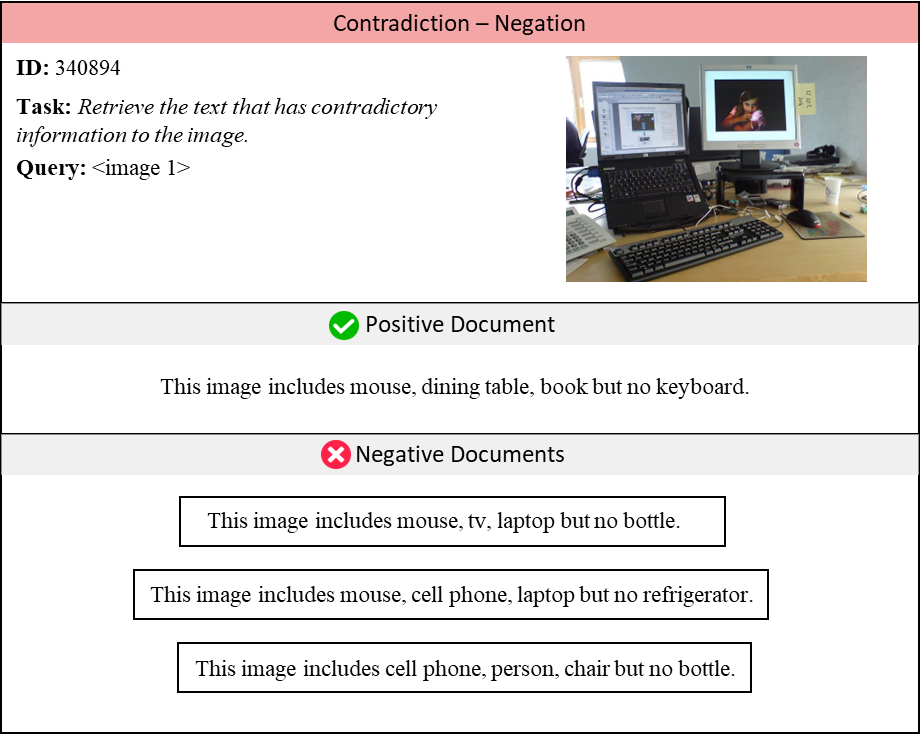}
    \caption{Negation example.
    }
    \label{case: negation}
\end{figure}

\begin{figure}[h]
    \centering
    \includegraphics[width=0.8\textwidth]{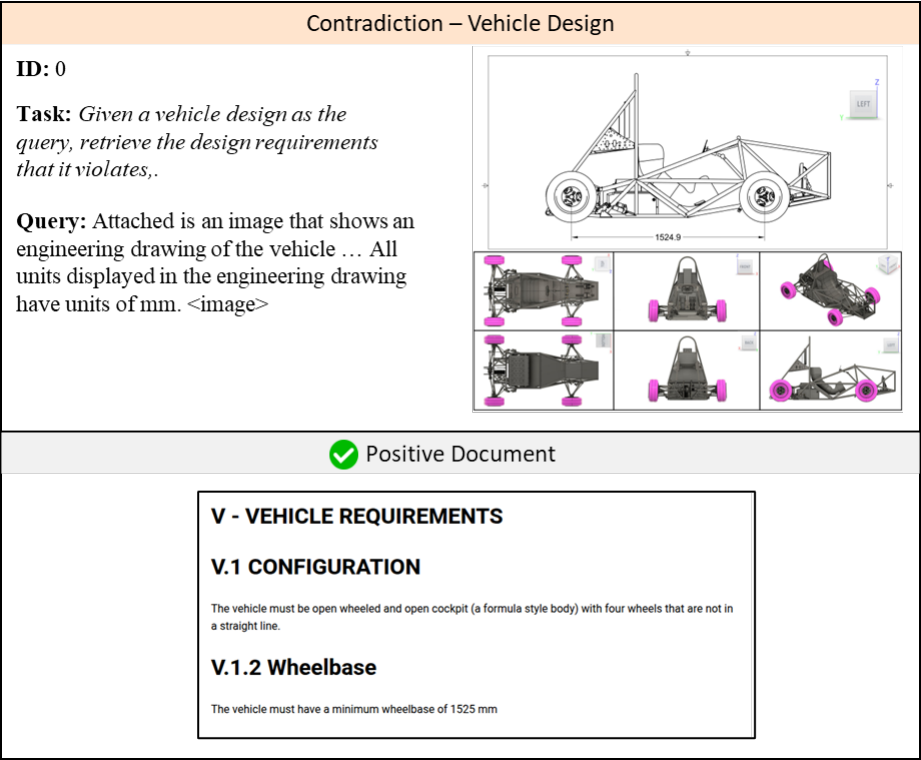}
    \caption{Vehicle Design example.
    }
    \label{case: design}
\end{figure}

\begin{figure}[h]
    \centering
    \includegraphics[width=0.8\textwidth]{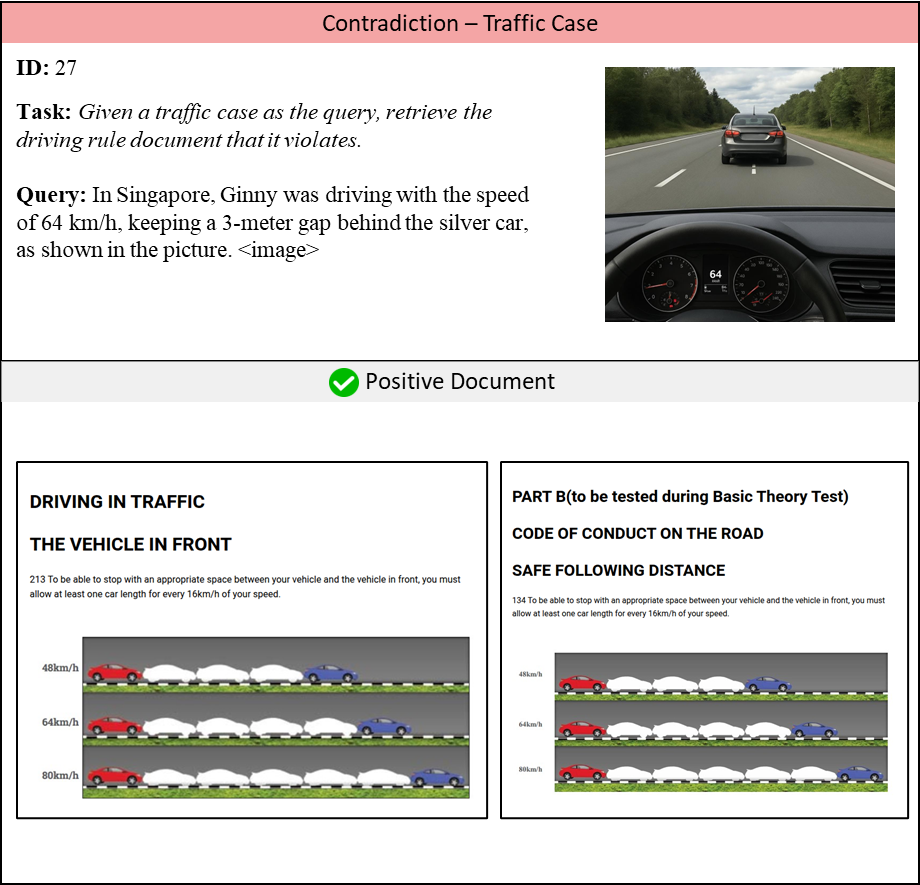}
    \caption{Traffic Case example.
    }
    \label{case: traffic}
\end{figure}


\end{document}